\pdfoutput=1
\documentclass[%
 reprint,
 superscriptaddress,
 selectp,
 amsmath,amssymb,
 aps,
 prapplied,
]{revtex4-1}
\usepackage{selectp}

\newcommand{\bfm}[1]{\mbox{\boldmath ${#1}$}}

\newcommand\eq[1] {(\ref{#1})} 
\newcommand{\beqa}{\begin{eqnarray}}
\newcommand{\eeqa}[1]{\label{#1}\end{eqnarray}}
\newcommand{\beq}{\begin{equation}}
\newcommand{\eeq}[1]{\label{#1}\end{equation}}

\newcommand{\Div}{\nabla \cdot}
\newcommand{\Curl}{\nabla \times}

\newcommand{\Tr}{\mathop{\rm Tr}\nolimits}

\newcommand{\lang}{\langle}
\newcommand{\rang}{\rangle}

\newcommand{\Ga}{\alpha}
\newcommand{\Gb}{\beta}

\newcommand{\Gve}{\varepsilon}

\newcommand{\Gc}{\chi}

\newcommand{\Gm}{\mu}

\newcommand{\Gz}{\zeta}

\newcommand{\BGve}{\bfm\varepsilon}

\newcommand{\bpm}{\begin{pmatrix}}
	\newcommand{\epm}{\end{pmatrix}}

\usepackage{url}
\usepackage{hyperref}

\newcommand\mydots{\ifmmode\ldots\else\makebox[1em][c]{.\hfil.\hfil.}\thinspace\fi}

\newcommand{\overbar}[1]{\mkern 1.5mu\overline{\mkern-1.5mu#1\mkern-1.5mu}\mkern 1.5mu}

\usepackage{graphicx}
\usepackage{dcolumn}
\usepackage{bm}
\usepackage{enumitem}

\usepackage{tikz}
\usetikzlibrary{arrows}

\tikzset{
	treenode/.style = {align=center, inner sep=0pt, text centered,
		font=\sffamily},
	arn_b/.style = {treenode, circle, blue, draw=blue, 
		text width=1.5em, very thick},
	arn_r/.style = {treenode, circle, red, draw=red, 
		text width=1.5em, very thick}
}

\begin{document}

\title{Tight Bounds on the Effective Complex Permittivity of Isotropic Composites\\ and Related Problems}

\author{Christian Kern}
 \email{kern@math.utah.edu}
 \affiliation{Department of Mathematics, University of Utah, Salt Lake City, UT 84112, USA}
\author{Owen D. Miller}
\affiliation{Department of Applied Physics and Energy Sciences Institute, Yale University, New Haven, CT 06511, USA}
\author{Graeme W. Milton}
\affiliation{Department of Mathematics, University of Utah, Salt Lake City, UT 84112, USA}

\begin{abstract}
Almost four decades ago, Bergman and Milton independently showed that the isotropic effective electric permittivity of a two-phase composite material with a given volume fraction is constrained to lie within lens-shaped regions in the complex plane that are bounded by two circular arcs. An implication of particular significance is a set of limits to the maximum and minimum absorption of an isotropic composite material at a given frequency. Here, after giving a short summary of the underlying theory, we show that the bound corresponding to one of the circular arcs is at least almost optimal by introducing a certain class of hierarchical laminates. In regard to the second arc, we show that a tighter bound can be derived using variational methods. This tighter bound is optimal as it corresponds to assemblages of doubly coated spheres, which can be easily approximated by more realistic microstructures. We briefly discuss the implications for related problems, including bounds on the complex polarizability.
\end{abstract}

\maketitle

\section{Introduction}
\setcounter{equation}{0}
Composite materials can exhibit effective ``metamaterial'' properties that are quite different from those of their underlying constituents \cite{Milton:2002:TOC, Tretyakov:2016:OMC, Kadic:2019:TDM} but the range of such emergent properties is not limitless. The effective permittivity of a two-phase isotropic composite with fixed volume fraction is constrained by the ``Bergman-Milton'' (BM) bounds \cite{Bergman:1980:ESM,Milton:1980:BCD, Milton:1981:BCP, Milton:1981:BTO,Bergman:1982:RBC} to lie within two circular arcs in the complex plane, yet the feasibility of approaching the extreme permittivity values---or, conversely finding tighter bounds---has remained largely an open question. In this paper, we resolve this question: we show that one of the arcs is nearly attainable via hierarchical laminates, we show that the other arc can be replaced by a tighter bound that can be derived by variational methods, and we identify assemblages of doubly coated spheres as structures that can achieve the extreme permittivity values at the new boundary. To design the hierarchical laminates that reach or approach the first arc, we start with five classes of microstructures that have previously been identified as optimal over narrow regions of BM bounds. Forming laminates from these microstructures leads to the identification of additional optimal microstructures and to broad coverage near the entirety of the first arc. To replace the second arc, we embed a rank-2 transformed permittivity tensor into a rank-4 effective tensor and use the ``Cherkaev-Gibiansky transformation'' \cite{Cherkaev:1994:VPC} together with the ``translation method'' \cite{Tartar:1979:ECH,Lurie:1982:AEC,Lurie:1984:EEC,Murat:1985:CVH,Tartar:1985:EFC} to maximally constrain the tensor values, adapting techniques used for related questions on the effective complex bulk modulus \cite{Gibiansky:1993:EVM}. These results provide a comprehensive understanding of possible effective metamaterial permittivities. For applications targeting extreme response, such as maximal absorption, these results refine the global bounds on what is possible and offer powerful microstructure design principles toward achieving them.

In the following, we study the effective-permittivity problem in the quasistatic regime for isotropic microstructures that are made from two isotropic materials (phases). We assume that the volume fractions $f_1$ and $f_2=1-f_1$ of the two phases are given, which is typically the problem of interest in applications in which the weight, or cost, of one of the phases is an issue. However, our results are not limited to fixed volume fractions, as the corresponding results for arbitrary volume fractions follow as a simple corollary.

The quasistatic regime corresponds to the assumption that the structures are periodic and that the
wavelength and attenuation lengths in the phases and the composite are much longer than the unit cell of periodicity. In this case, one can use the quasistatic equations (see Sec.\,11.1 in Ref.\,\onlinecite{Milton:2002:TOC})
\beq \bm{d}=\Gve\,\bm{e},\quad \Curl\bm{e}=0, \quad \Div\bm{d}=0,\quad \Gve=\Gc\Gve_1+(1-\Gc)\Gve_2, \eeq{0.a}
where $\bm{d}(\bm{x})$ and $\bm{e}(\bm{x})$ are complex periodic vector fields, with the real parts of $\bm{d}(\bm{x})\text{e}^{-\text{i}\omega t}$ and $\bm{e}(\bm{x})\text{e}^{-\text{i}\omega t}$ representing the electric displacement field and the electric field, respectively, $\Gc(\bm{x})$ is the periodic characteristic
function that takes the value $1$ in phase $1$ and $0$ in phase $2$, and $\Gve_1$ and $\Gve_2$ are the (frequency-dependent) complex electric
permittivities of the two materials. Letting $\lang \cdot\rang$ denote a volume average over the unit cell of periodicity, we may solve these equations
for any value of $\lang \bm{e}\rang$ (provided $\Gve_1/\Gve_2$ is finite, nonzero, and non-negative). Then, since $\lang \bm{d}\rang$ depends linearly on $\lang \bm{e}\rang$,
we may write
\beq \lang\bm{d}\rang=\Gve_*\lang\bm{e}\rang, \eeq{0.b}
which defines the effective complex electric permittivity, $\Gve_*$.

The goal is then to find the range that $\Gve_*$ takes as the geometry, i.e., $\Gc(\bm{x})$, varies over all periodic configurations with an isotropic effective permittivity that have the prescribed
volume fraction $f_1$ of phase $1$. The geometries that realize (or almost realize) the maximum and minimum values of the imaginary part of $\Gve_*$  are those
that show (close to) the maximum and minimum amount of absorption of an incoming plane wave or a constant static applied field. It is not only the imaginary part of $\Gve_*$ that is of interest, as both the real and imaginary parts
are of importance in determining the effective refractive index and the transmission and reflection at an interface. If there is a slow
variation (relative to the local periodicity) in the microstructure, the resulting variations in $\Gve_*(\bm{x})$ can be used to guide waves.  

Substantial progress on this problem was made about four decades ago when bounds on the effective complex electric permittivity were derived
\cite{Bergman:1980:ESM,Bergman:1982:RBC,Milton:1980:BCD, Milton:1981:BCP, Milton:1981:BTO} (see also Chaps. 18, 27, and 28 in Ref.~\onlinecite{Milton:2002:TOC}), which have become known as the ``Bergman-Milton'' (BM) bounds. The BM bounds comprise bounds for several different restrictions on the geometry of the structure, including the problem we are studying here, i.e., bounds for structures with isotropic effective permittivity (which includes, for example, all geometries with cubic symmetry) and fixed volume fractions.

Originally, the BM bounds were derived on the basis of the analytic properties of $\Gve_*$ as a function of $\Gve_1$ and $\Gve_2$. Bergman, in his pioneering work \cite{Bergman:1978:DCC}, had recognized the analytic properties but erroneously assumed that the function would be rational for any periodic geometry (a checkerboard is a counterexample \cite{Dykhne:1970:CTD, Milton:1979:TST}; see also Chap.\,3 in Ref.~\onlinecite{Milton:2002:TOC}) and that if the equations did not have a solution for a prescribed average electric field, then they would have a solution for a prescribed average displacement field (a square array of circular disks having $\Gve_1=-1$ inside the disks and $\Gve_2=1$ outside the disks is a counterexample in which neither has a solution) \cite{Milton:1979:TST}. An argument that avoided these difficulties by approximating the composite by a discrete network of two electric impedances has been put forward \cite{Milton:1981:BCP} and, later, the analytic properties have been rigorously established by Golden and Papanicolaou \cite{Golden:1983:BEP}. 

The BM bounds can also be used in an inverse fashion, i.e., to obtain information about the composition of a material from a measurement of its effective properties. If we know that a composite material is made from two phases with known permittivities, we can use the BM bounds to obtain bounds on the volume fraction from a measurement of its effective permittivity \cite{McPhedran:1990:ITP,Cherkaeva:1998:IBM}. In essence, one has to find the range of values of the volume fraction for which the measured value of the effective permittivity lies in the lens-shaped region. 

We remark, in passing, that the analytic method is also useful for bounding the response in the time domain for a given time-dependent applied field
(that is not at constant frequency) \cite{Mattei:2016:BRV}. This approach is more useful for the equivalent antiplane elasticity problem, since typical relaxation times are much longer.

Furthermore, the analytic properties extend to the Dirichlet-to-Neumann map governing the response of bodies containing two or more phases, not just in quasistatics but also for wave equations (at constant frequency) (see Chaps.\,3 and 4 in \cite{Milton:2016:ETC}).

Before discussing the BM bounds in more detail, we briefly consider the analytic properties of the electric permittivity as a function of frequency \cite{Landau:1984:ECMa,Jackson:1999:CE}, which are a result of the fundamental restrictions imposed by causality and passivity. Causality, i.e., the fact that the electric displacement field at a certain point in time depends on the electric field at prior times only, implies that the frequency-dependent permittivity, $\varepsilon(\omega)$, is an analytic function in the upper halfplane. If additionally, the material is passive, i.e., if it does not produce energy, the imaginary part of the permittivity is non-negative for positive real frequencies. These properties, which in mathematical terms mean that the permittivity can be expressed via a Stieltjes or a Nevanlinna/Herglotz function \cite{Baker:1981:PAB,Akhiezer:1993:LOH,Bernland:2011:SRC,Cassier:2017:BHF}, have far-reaching and not immediately intuitive consequences. The Kramers-Kronig relations, which express the real (dispersive) part of the permittivity in terms of its imaginary (absorptive) part and vice versa, constitute a well-known example. Moreover, the analytic properties lead to sum rules, i.e, relations connecting integral quantities involving the permittivity with its static and high-frequency behavior, which prove useful in a variety of applications. For example, sum rules have been used to derive bounds on broadband cloaking in the quasistatic regime \cite{Cassier:2017:BHF} and on dispersion in metamaterials \cite{Gustafsson:2010:SRP}. As one may expect, such considerations are not limited to electromagnetics but apply to transfer functions of passive linear time-invariant (LTI) systems in general \cite{Bernland:2011:SRC}.

In order to derive the BM bounds, one uses the fact that the analytic properties extend to the permittivity as a function of the constituent phases \cite{Golden:1983:BEP}. More precisely, one uses the fact that $\Gve_*(\Gve_1,\Gve_2)$ is a homogeneous function of $\Gve_1$ and $\Gve_2$ (so it suffices to consider the function with $\Gve_2=1$)
and $f(z)=\Gve_*(1/z,1)$ is a Stieltjes function of $z$, thus having the integral representation
\beq f(z)=\Ga+\frac{\Gb}{z}+\int_0^\infty\frac{\textrm{d}\Gm(t)}{t+z}, \eeq{0.0}
for all $z$ not on the negative real axis of the complex plane, where $\Ga$ and $\Gb$ are non-negative real constants and $\textrm{d}\Gm$ is a non-negative measure on $(0,\infty)$. Depending on the assumptions about the geometry of the composite, the Stieltjes function satisfies different constraints \cite{Bergman:1978:DCC, Golden:1983:BEP}: First, for an arbitrary, potentially anisotropic composite material, the Stieltjes function satisfies
\begin{eqnarray} 
f(1)=1.
\end{eqnarray}
Second, if the volume fraction of phase 1 in the composite, $f_1$, is prescribed, the Stieltjes function additionally satisfies
\begin{eqnarray}  
\frac{\textrm{d}f(z)}{\textrm{d}z}{\bigg|}_{z=1}=-f_1. 
\end{eqnarray}
Third, if the composite is assumed to be isotropic, the Stieltjes function additionally satisfies
\begin{eqnarray} 
\frac{\textrm{d}^2f(z)}{\textrm{d}z^2}{\bigg|}_{z=1}=2f_1-\frac{2f_1f_2}{3}.
\end{eqnarray}
The goal is then to determine the sets of values that the function $f(z)$ and, hence, the effective permittivity (or any diagonal element of the effective permittivity tensor) attains as the geometry, i.e., the measure $\textrm{d}\mu$, is varied while being subject to the first constraint (anisotropic composites), the first and the second constraint (anisotropic composites with fixed volume fraction) and all of the constraints (isotropic composites with fixed volume fraction). 

It can be shown that mappings between these sets are realized by linear fractional transformations \cite{Golden:1983:BEP,Bergman:1993:HSF} (see also Chap.\,28 in \cite{Milton:2002:TOC}). As these transformations map generalized circles (circles or straight lines) onto generalized circles, one obtains a set of nested bounds, the BM bounds, that confine the effective permittivity to lens-shaped regions bounded by circular arcs. It should be pointed out that in the larger arena of Stieltjes functions, lens-shaped bounds corresponding to the BM bounds and their generalizations have a long history (see, e.g., Refs.\,\onlinecite{Krein:1938:CNP,Henrici:1966:TEE,Gragg:1968:TEB,Krein:1977:MMP}).

\begin{figure}
	\includegraphics{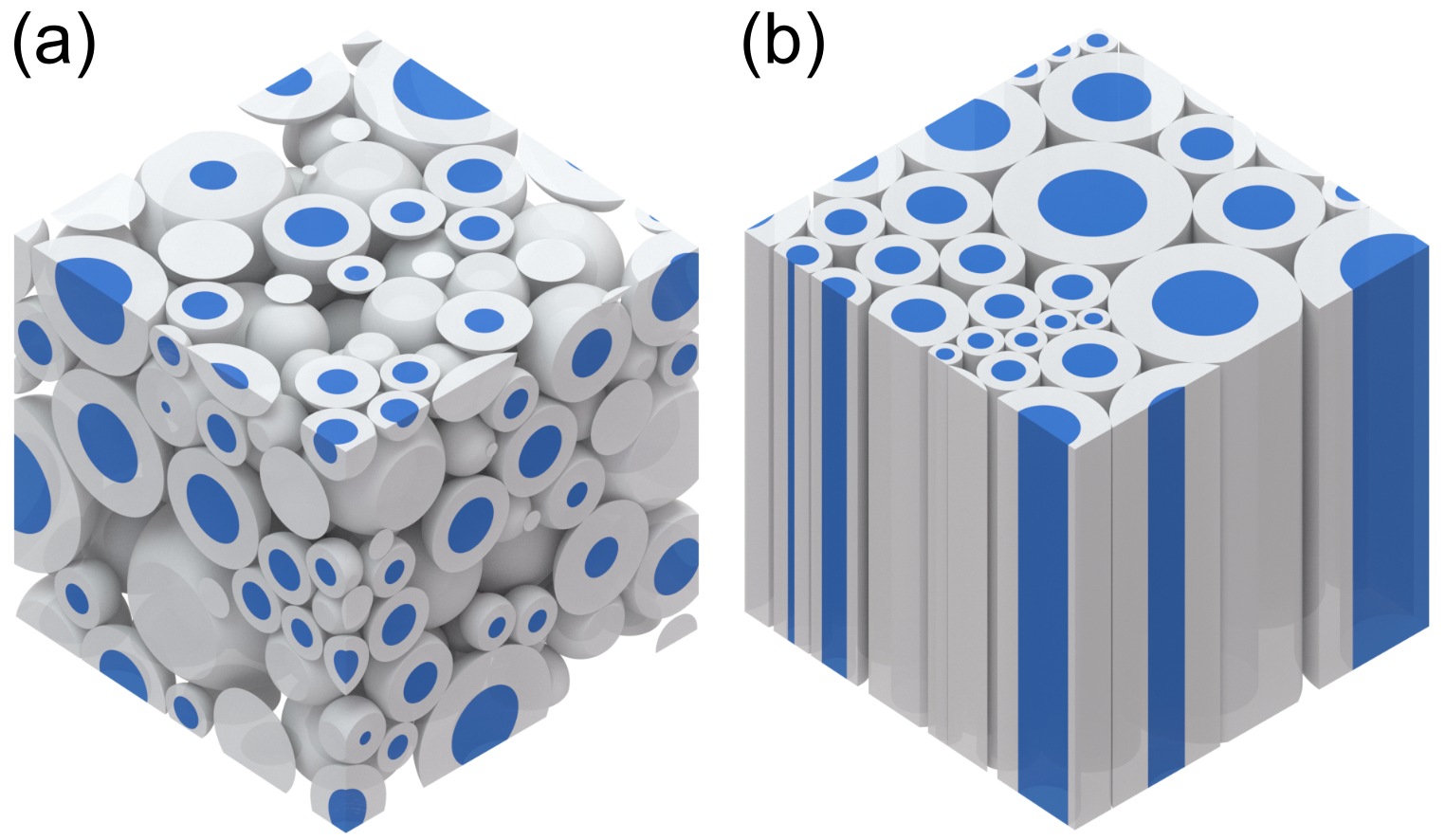}
	\caption{Schematic illustrations of the Hashin-Shtrikman assemblages of coated spheres (a) and coated cylinders (b). The spheres (cylinders) are identical except for a scaling factor and fill all space.}
	\label{fig1}
\end{figure}

The BM bounds for an isotropic composite material with fixed volume fraction confine $\Gve_*$ to lie within the following lens-shaped region in the complex plane. One side is the circular arc traced by
\beq \Gve_*^+(u_1)=f_1\Gve_1+f_2\Gve_2-\frac{f_1f_2(\Gve_1-\Gve_2)^2}{3(u_1\Gve_1+u_2\Gve_2)}, \eeq{0.1}
as $u_1$ is varied so that $u_1\geq f_2/3$ and $u_2\geq f_1/3$ while keeping $u_1+u_2=1$ \cite{Milton:1981:BCP}. The essential
feature of this formula is that it has only one pole (resonance) at finite negative values of the ratio $\Gve_1/\Gve_2$. On the other side
is the circular arc traced by
\beq \Gve_*^-(u_1)=\left(\frac{f_1}{\Gve_1}+\frac{f_2}{\Gve_2}-\frac{2f_1f_2(1/\Gve_1-1/\Gve_2)^2}{3(u_1/\Gve_1+u_2/\Gve_2)}\right)^{-1}, \eeq{0.2}
as $u_1$ is varied so that $u_1\geq 2f_2/3$ and $u_2\geq 2f_1/3$ while keeping $u_1+u_2=1$ \cite{Milton:1981:BCP}. The two circular arcs
meet at the points
\begin{eqnarray}  \Gve_*^+(f_2/3)&& = \Gve_*^-(2f_2/3)=\Gve_2+\frac{3f_1\Gve_2(\Gve_1-\Gve_2)}{3\Gve_2+f_2(\Gve_1-\Gve_2)}, \label{0.3}\\
\Gve_*^+(1-f_1/3)&& = \Gve_*^-(1-2f_1/3)=\Gve_1+\frac{3f_2\Gve_1(\Gve_2-\Gve_1)}{3\Gve_1+f_1(\Gve_2-\Gve_1)}. \nonumber
\end{eqnarray}
When $\Gve_1$ and $\Gve_2$ are real, the lens-shaped region collapses to an interval between these two points, thus giving the well-known Hashin-Shtrikman bounds
\cite{Hashin:1962:VAT}.
For this reason the BM bounds have sometimes been called the complex Hashin-Shtrikman bounds, which might be considered an occurrence of Stigler's law \cite{Stigler:1980:SLE}, as Hashin and Shtrikman had
nothing to do with their derivation. The two points given in Eq.\,\eq{0.3} correspond to the Hashin-Shtrikman assemblages of coated spheres that fill all space, each being identical to one another, apart from
a scale factor \cite{Hashin:1962:VAT}. Illustrations of such a coated-sphere assemblage and its columnar counterpart, the coated-cylinder assemblage, are shown in Fig\,\ref{fig1}.

\begin{figure}
	\includegraphics{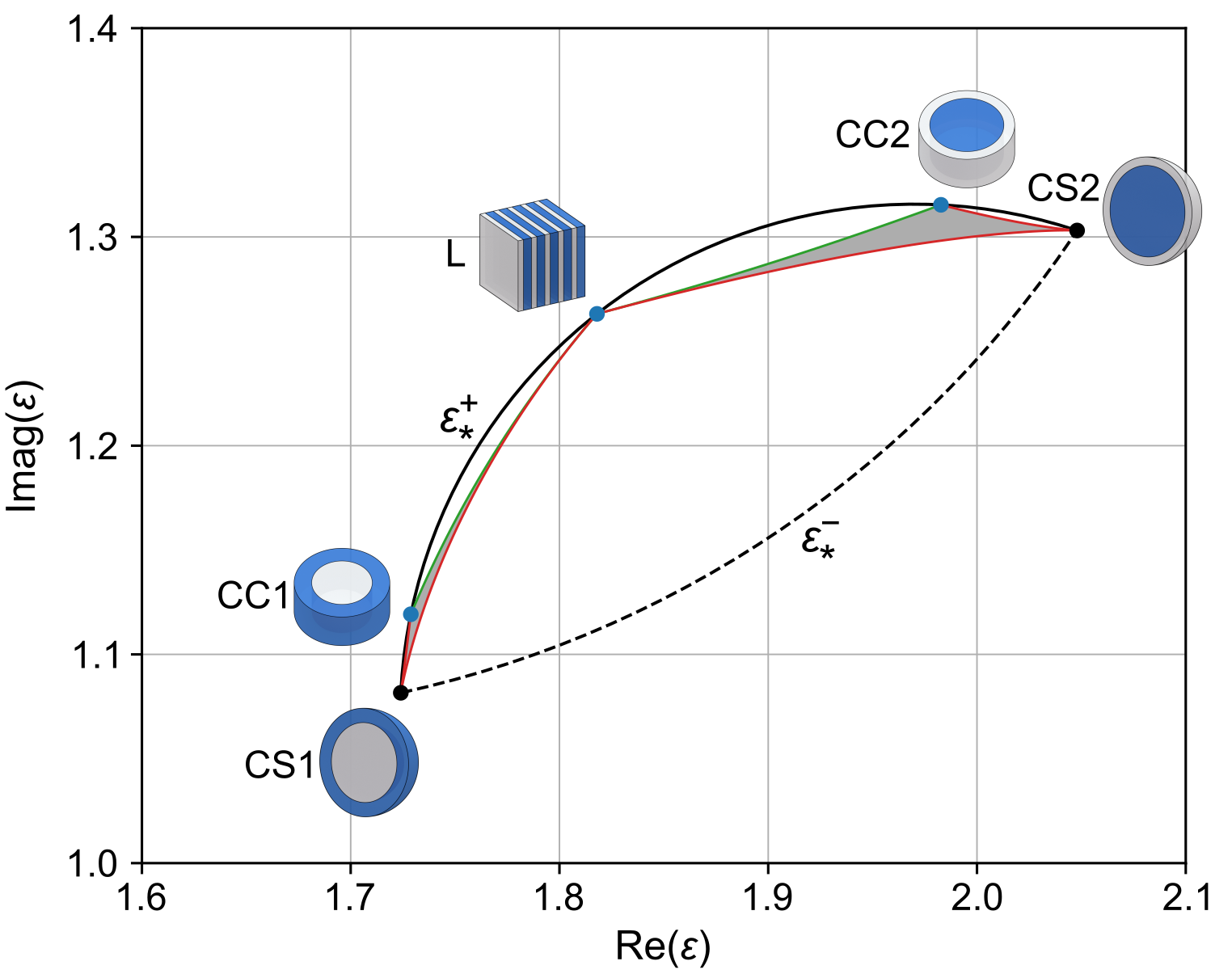}
	\caption{An illustration of the Bergman-Milton bounds for an isotropic composite material with fixed volume fraction. The effective complex permittivity is constrained to a lens-shaped region bounded by the two circular arcs, $\Gve_*^+(u_1)$ and $\Gve_*^-(u_1)$. The five points on the arc $\Gve_*^+(u_1)$ correspond to the two Hashin-Shtrikman assemblages of coated spheres with phase $1$ and phase $2$ as the core material, CS1 and CS2, respectively, Schulgasser laminates of the two Hashin-Shtrikman assemblages of coated cylinders, CC1 and CC2, and a Schulgasser laminate formed from a simple rank-$1$ laminate, L. The gray-shaded region corresponds to Schulgasser laminates formed from assemblages of coated ellipsoids. It is bounded by Schulgasser laminates of coated elliptical cylinders and coated spheroids, corresponding to the green and red curves, respectively. The parameters are $\varepsilon_1 = 0.2+1.5\text{i}$, $\varepsilon_2 = 3+0.4\text{i}$, and $f_1=0.4$.}
	\label{fig2}
\end{figure}

In Ref.~\onlinecite{Milton:1981:BCP}, Milton identified microstructures that attain three additional points on the arc $\Gve_*^+(u_1)$. His key idea was to look for microstructures that have only one pole, as this is the characteristic feature of the bound \eq{0.1}. If one finds such a microstrucure with a diagonal anisotropic effective tensor $\BGve_*$, then, by forming a so-called Schulgasser laminate \cite{Schulgasser:1977:BCS}, one can obtain a composite with isotropic effective permittivity $\Tr(\BGve_*)/3$. As described in detail in Sec.\,22.2 of Ref.\,\onlinecite{Milton:2002:TOC}, the corresponding lamination scheme is based on the following observation. Consider two materials with diagonal permittivity tensors and the same principal permittivity in one direction, such as when one is a $90^{\circ}$ rotation of the other about this direction. If these two materials are laminated along this direction, then the effective permittivity tensor of the laminate is just an arithmetic mean of the permittivity tensors of the two materials. By successively applying this idea on widely separated length scales, starting from the anisotropic material with diagonal effective tensor $\bm{\varepsilon}_*$, one obtains a hierarchical laminate with isotropic effective permittivity $\Tr(\BGve_*)/3$. If one takes $\BGve_*$ to be the effective permittivity of a simple laminate, an assemblage of coated cylinders with a core of phase $1$ and a coating of phase $2$, and an assemblage of coated cylinders with a core of phase $2$ and a coating of phase $1$, this results in the effective permittivities $\Gve_*^+(f_2)$, $\Gve_*^+(f_2/2)$, and $\Gve_*^+(1-f_1/2)$, respectively, which, as illustrated in Fig.\,\ref{fig2}, all lie on the arc $\Gve_*^+(u_1)$. 

These three microstructures as well as the Hashin-Shtrikman coated-sphere assemblage can be seen as special cases of Schulgasser laminates formed from assemblages of coated ellipsoids (those having only one pole). The effective permittivity tensor, $\bm{\varepsilon}_*$, of such assemblages can be obtained from the following formula \cite{Milton:2002:TOC},
\beq f_1 \varepsilon_1(\bm{\varepsilon}_*-\varepsilon_2\bm{I})^{-1}=\varepsilon_2(\varepsilon_1-\varepsilon_2)^{-1}\bm{I}+f_2\bm{M},
\eeq{0.5}
where $\bm{I}$ is the identity matrix and $\bm{M}$ is a matrix that depends on the depolarization tensors of the ellipsoids. As the shape and orientation of the ellipsoids is varied, $\bm{M}$ ranges over all positive-semidefinite symmetric matrices with $\text{Tr}(\bm{M})=1$ (see Sec.\,7.8 in Ref.~\onlinecite{Milton:2002:TOC}). If one chooses the coordinate system such that its axes coincide with the principal axes of the ellipsoids, $\bm{M}$ is a diagonal matrix. The range of effective permittivities of Schulgasser laminates formed from such assemblages is illustrated in Fig.\,2. It is bounded by Schulgasser laminates formed from assemblages of coated elliptical cylinders, $\bm{M}=(\alpha, 1-\alpha, 0)$ with $\alpha \in [1/2,1]$, and coated spheroids, $\bm{M}=(\alpha, \alpha, 1-2\alpha)$ with $\alpha=[0, 1/2]$.

The first main thrust of this paper is to show that there are hierarchical laminate geometries that attain, depending on the volume fraction, two or three more points on the arc $\Gve_*^+(u_1)$ and other hierarchical laminates, which typically come extremely close to attaining the entire arc. This result shows that any improved bound,
if it exists, typically can only be marginally better than Eq.\,\eq{0.1}. Note that while we are focussing on bounds on the complex permittivity, similar conclusions almost
certainly apply to the attainability of four related bounds: those coupling the two effective permittivities $\Gve_*=\Gve_*(\Gve_1,\Gve_2)$ and $\widetilde{\Gve}_*=\Gve_*(\widetilde{\Gve}_1,\widetilde{\Gve}_2)$ for given real positive values of $\Gve_1$, $\Gve_2$, $\widetilde{\Gve}_1$, $\widetilde{\Gve}_2$
\cite{Bergman:1978:DCC}; those associated bounds of Beran \cite{Beran:1965:UVA}
in the form simplified by Milton \cite{Milton:1981:BEE} and Torquato and Stell \cite{Torquato:1980:MAT,Torquato:1985:BET} (that can be obtained by taking the limit as
$\widetilde{\Gve}_1\to\widetilde{\Gve}_2\to 1$; see Ref.~\onlinecite{Avellaneda:1988:ECP})
that correlate, at fixed volume fraction, $\BGve_*$ with a geometric parameter $\Gz_1$, or $\Gz_2=1-\Gz_1$, that can be calculated from the three-point correlation function giving the probability that a triangle
positioned and oriented randomly in the composite has all three vertices in phase $1$, or respectively phase $2$; those bounds on the complex effective bulk modulus \cite{Gibiansky:1993:EVM}
of an isotropic composite of two isotropic elastic phases; and those bounds that couple the effective conductivity with the effective bulk modulus \cite{Gibiansky:1996:CBC} when the conductivities and
elastic moduli of the phases are real.
For all of these problems, the same five microstructures attaining the points $\Gve_*^+(f_2/3)$, $\Gve_*^+(f_2)$, $\Gve_*^+(f_2/2)$, $\Gve_*^+(1-f_1/2)$ and $\Gve_*^+(1-f_1/3)$
were also shown to attain the relevant bounds \cite{Milton:1981:BTO,Avellaneda:1988:ECP,Gibiansky:1993:EVM,Gibiansky:1996:CBC}.

It was noted in Ref.~\onlinecite{Milton:1981:BCP} that the formula \eq{0.2} does not satisfy the phase interchange inequality
\beq \Gve_*(\Gve_1,\Gve_2)\Gve_*(\Gve_2,\Gve_1)\geq\Gve_1\Gve_2 \eeq{0.4}
of Schulgasser \cite{Schulgasser:1976:PIR}, which holds when $\Gve_1$ and $\Gve_2$ are real and non-negative. Consequently, it has been suggested
that the bound \eq{0.2} is non-optimal \cite{Milton:1981:BCP}. In fact, using this inequality, an improved bound was obtained by Bergman \cite{Bergman:1982:RBC}. However, the inequality is itself nonoptimal and a tighter inequality,
\beq \frac{\Gve_*(\Gve_1,\Gve_2)\Gve_*(\Gve_2,\Gve_1)}{\Gve_1\Gve_2}+\frac{\Gve_*(\Gve_1,\Gve_2)+\Gve_*(\Gve_2,\Gve_1)}{\Gve_1+\Gve_2}\geq 2,
\eeq{0.5}
has been proposed \cite{Milton:1981:BCP}, and partially proved \cite{Avellaneda:1988:ECP} with an error in the proof corrected
in Refs.~\onlinecite{Nesi:1991:MII,Zhikov:1991:EHM}. As remarked in
Ref.~\onlinecite{Milton:1981:BCP}, this inequality holds as an equality for any assemblage of multicoated spheres where all multicoated spheres in the assemblage
are identical apart from a scale factor. In two dimensions, the Schulgasser inequality, Eq.\,\eq{0.4},
holds as an equality not just for coated-disk assemblages, but for any geometry in which the (two-dimensonal) effective permittivity is isotropic
\cite{Keller:1964:TCC}. The identity can be used to improve the bounds on the complex permittivity for isotropic two-dimensional composites of two isotropic
phases, or equivalently the transverse conductivity of a three-dimensional geometry where the conductivity does not vary in the axial direction, and the
resulting bounds \cite{Milton:1980:BCD, Milton:1981:BCP} are attained by assemblages of doubly coated disks. (The claim in Ref.~\onlinecite{Bergman:1980:ESM}
that these bounds are wrong was unfounded.) This suggests that the doubly-coated-sphere assemblage (with the appropriate phase at the central core) may in fact correspond to an optimal bound on the effective complex permittivity, replacing the bound \eq{0.2}. Additional evidence is that for the four related bounding problems mentioned above, the doubly-coated-sphere assemblages make their appearance in attaining a bounding curve.

The second main thrust of this paper is to derive this improved bound, replacing Eq.\,\eq{0.2}, utilizing minimization variational principles for the complex effective permittivity derived by
Gibiansky and Cherkaev \cite{Cherkaev:1994:VPC} (that also apply to other problems with complex moduli, such as viscoelasticity \cite{Cherkaev:1994:VPC},
and which have later been extended to other non-self-adjoint problems \cite{Milton:1990:CSP}, to wave equations in lossy media \cite{Milton:2009:MVP,Milton:2010:MVP}, and to scattering
problems \cite{Milton:2017:BCP}). These variational principles allow one to use powerful techniques for deriving bounds, namely the variational approach of
Hashin and Shtrikman \cite{Hashin:1962:VAT} and the translation method, also known as the method of compensated compactness,
of Tartar and Murat \cite{Tartar:1979:ECH,Murat:1985:CVH,Tartar:1985:EFC} and
Lurie and Cherkaev \cite{Lurie:1982:AEC,Lurie:1984:EEC} (see also Refs.\, \cite{Torquato:2001:RHM,Cherkaev:2000:VMS,Milton:2002:TOC,Tartar:2009:GTH}). One advantage of the variational approach, as opposed to the analytic approach of Bergman and Milton, is that it easily extends to multiphase media and to media with anisotropic phases (including polycrystalline media) that have complex permittivity tensors or complex elasticity tensors. Some elementary bounds on the effective complex permittivity tensor and effective complex elasticity tensor are given in Ref.~\onlinecite{Milton:1990:CSP} (see also Sec.\,22.6 in Ref.~\cite{Milton:2002:TOC}). Some of these bounds have also been conjectured or derived using the analytic approach \cite{Golden:1985:BEP,Golden:1986:BCP,Bergman:1986:EDC,Milton:1987:MCEa,Milton:1987:MCEb,Milton:1990:RCF,Gully:2015:BCP} but generally with much greater difficulty. Our analysis is close to that used in Ref.~\onlinecite{Gibiansky:1993:EVM} to derive bounds on the effective complex bulk modulus, which has later been extended
to bounds on the effective complex shear modulus \cite{Milton:1997:EVM,Gibiansky:1999:EVM} (see also Refs.~\onlinecite{Gibiansky:1993:BCB,Gurevich:2017:OBA}).

The BM bound and our improved bound, like many bounds on effective moduli that involve the volume fractions of the phases (as well as possibly other information)
simplify when expressed in terms of their $y$-transforms (see Chaps.\,19 and 26 and Secs.\,23.6 and 24.10 in Ref.~\onlinecite{Milton:2002:TOC} and references therein).
Rather than working with bounds on $\Gve_*$, one works with bounds on its $y$-transform
\beq y_\Gve=-f_2\Gve_1-f_1\Gve_2+\frac{f_1f_2(\Gve_1-\Gve_2)^2}{f_1\Gve_1+f_2\Gve_2-\Gve_*}, \eeq{0.6}
in terms of which
\beq \Gve_*=f_1\Gve_1+f_2\Gve_2-\frac{f_1f_2(\Gve_1-\Gve_2)^2}{f_2\Gve_1+f_1\Gve_2+y_\Gve}. \eeq{0.7}
The BM bounds confine $y_\Gve$ to a volume-fraction-independent lens-shaped region bounded by the straight line joining $2\Gve_1$ and $2\Gve_2$, corresponding to the bound \eq{0.1}, and a segment
of a circular arc joining $2\Gve_1$ and $2\Gve_2$ that, when extended, passes through the origin, corresponding to the bound \eq{0.2}. Our improved bound, which replaces this circular arc, is the outermost of two circular arcs joining  $2\Gve_1$ and $2\Gve_2$: one arc when extended passes through $-\Gve_1$, while the other arc when extended passes through $-\Gve_2$.

Note that while we obtain an improved bound for isotropic composites, our bound also applies to anisotropic composites if we replace the scalar effective permittivity, $\varepsilon_*$, with $\text{Tr}(\bm{\varepsilon}_*)/3$ or any other effective permittivity of isotropic polycrystals.

\section{On the optimality of the single-resonance bound}
In the following, we show that the bound \eq{0.1} is at least almost optimal. First, we derive expressions for the $y$-transformed effective permittivities of the five known optimal microstructures and discuss corresponding hierarchical laminates. We then show that there are, depending on the volume fraction, at least two or three additional optimal laminates. In the last part of this section, using numerical calculations, we consider related laminates that come very close to attaining the bound.

The first two of the five microstructures that are known to attain the bound \eq{0.1} are the two Hashin-Shtrikman coated-sphere assemblages (CS). For phase $1$ as the core material, the $y$-transformed effective permittivity of the assemblage is given by $y_{\varepsilon}^{\text{CS1}}=2\varepsilon_2$. For phase $2$ as the core material, one obtains $y_{\varepsilon}^{\text{CS2}}=2\varepsilon_1$. The third and the fourth optimal microstructure are Schulgasser laminates formed from assemblages of coated cylinders (CC). The effective permittivity of such an assemblage in the directions perpendicular to the cylinder axes is given by
\beq \varepsilon_*^{\text{CC}1,\perp}=\varepsilon_2+\frac{2f_1\varepsilon_2(\varepsilon_1-\varepsilon_2)}{2\varepsilon_2+f_2(\varepsilon_1-\varepsilon_2)},
\eeq{1.1}
while parallel to the cylinder axes, one obtains
\beq \varepsilon_*^{\text{CC}1,\parallel}=f_1\varepsilon_1+f_2\varepsilon_2.
\eeq{1.2}
The corresponding Schulgasser laminate has effective permittivity
\beq \varepsilon_*^{\text{CC}1}=\frac{1}{3}\left(2\varepsilon_*^{\text{CC}1,\perp}+\varepsilon_*^{\text{CC}1,\parallel}\right)
\eeq{1.3}
and its $y$-transformed effective permittivity is given by
\beq y_{\varepsilon}^{\text{CC}1}=f_1y_{\varepsilon}^{\text{CS}1}+f_2\left(\frac{3}{4}y_{\varepsilon}^{\text{CS}1}+\frac{1}{4}y_{\varepsilon}^{\text{CS}2}\right).
\eeq{1.4}
Analogously, one obtains for the phase-interchanged microstructure, i.e., for phase 2 as the core material,
\beq y_{\varepsilon}^{\text{CC}2}=f_2y_{\varepsilon}^{\text{CS}2}+f_1\left(\frac{3}{4}y_{\varepsilon}^{\text{CS}2}+\frac{1}{4}y_{\varepsilon}^{\text{CS}1}\right).
\eeq{1.5}
The fifth optimal microstructure, the Schulgasser laminate formed from a rank-1 laminate (L), has effective permittivity
\beq \varepsilon_*^{\text{L}}=\frac{1}{3}\left(2\left(f_1\varepsilon_1+f_2\varepsilon_2\right)+\left(\frac{f_1}{\varepsilon_1}+\frac{f_2}{\varepsilon_2}\right)^{-1}\right)
\eeq{1.6}
and, therefore,
\beq y_{\varepsilon}^{\text{L}}=f_1y_{\varepsilon}^{\text{CS}1}+f_2y_{\varepsilon}^{\text{CS2}}=f_1\cdot2\varepsilon_2+f_2\cdot2\varepsilon_1.
\eeq{1.7}
This last relation implies that every point on the $y$-transformed version of the bound \eq{0.1} is attained by a rank-1 laminate with some volume fraction $f_1\in\left[0,1\right]$. Hence, there can be no tighter bound that is volume-fraction independent when expressed in terms of the $y$-transformed complex permittivity. 

It is well known that for each assemblage of coated ellipsoids, there is a hierarchical laminate with the same effective permittivity (see Ref.~\onlinecite{Tartar:1985:EFC} and Chap.\,9 in Ref.~\onlinecite{Milton:2002:TOC}). Hierarchical laminates are laminates that are formed in more than one lamination step, i.e., they are laminates of laminates. It is assumed that the length scales of subsequent lamination steps, the number of which is referred to as the rank of the laminate, are sufficiently separated. More precisely, the coated-ellipsoid assemblages correspond to a specific type of hierarchical laminates, so-called coated laminates---first studied by Maxwell \cite{Maxwell:1954:TEM}---that are formed as follows. In the first lamination step, a laminate is formed from the two pure phases. One of these phases is referred to as the core phase, while the other phase is the so-called coating phase. In all subsequent lamination steps, the laminate obtained in the previous lamination step is laminated with the coating phase. For mutually orthogonal lamination directions and a particular choice of the volume fractions (see Chap.\,9 in Ref.~\onlinecite{Milton:2002:TOC}), one obtains rank-$2$ and rank-$3$ coated laminates equivalent to the Hashin-Shtrikman assemblages of coated cylinders and spheres, respectively. Thus, the five microstructures that are known to attain the bound \eq{0.1} can equivalently be seen as hierarchical laminates.

\begin{figure}
	\includegraphics{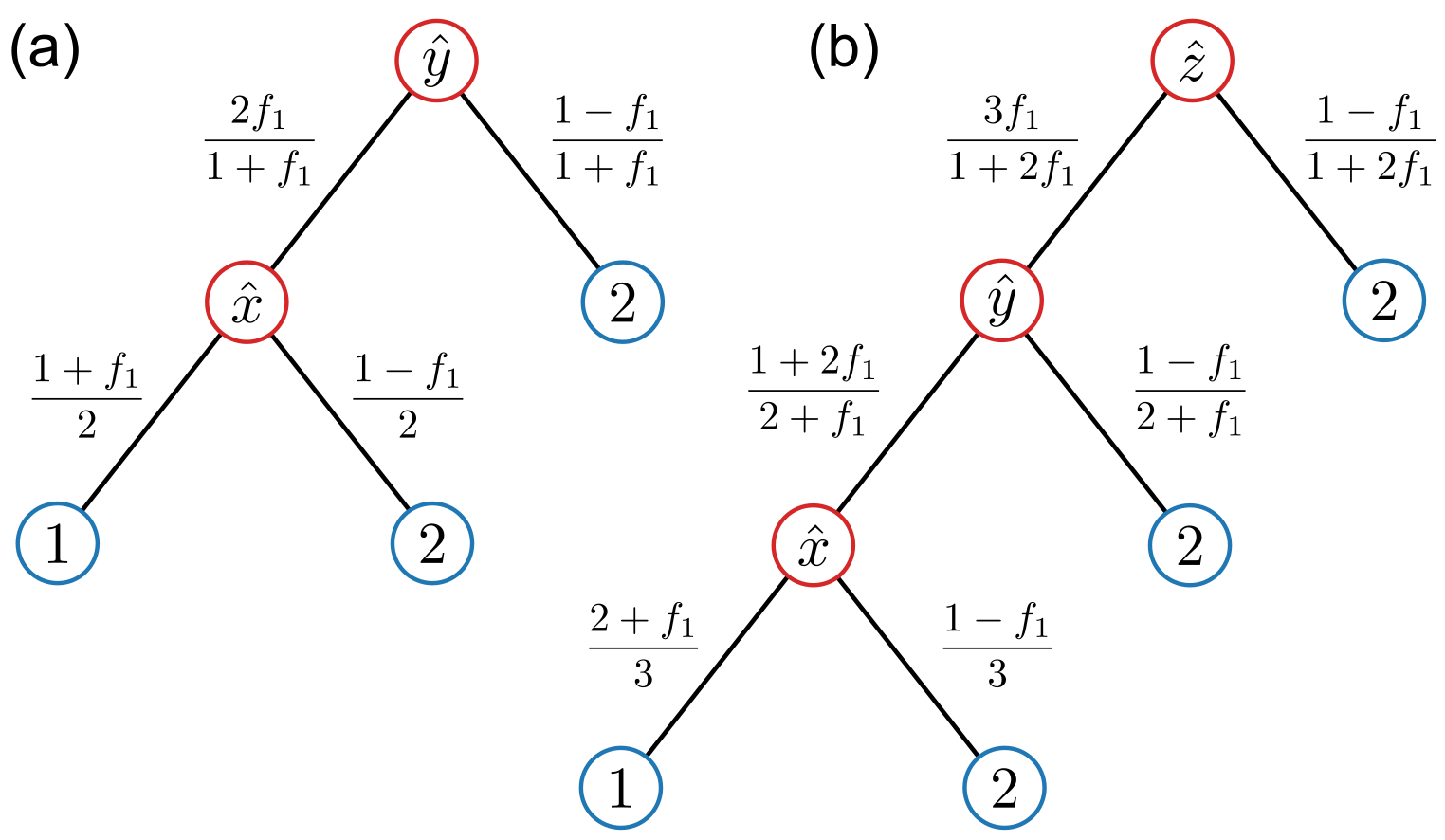}
	\caption{Tree structures corresponding to hierarchical laminates that have the same effective permittivity as the Hashin-Shtrikman coated-cylinder assemblage (a) and coated-sphere assemblage (b). The blue nodes, i.e., the leaves of the tree, correspond to one of the two pure phases, phase $1$ or phase $2$, while the red nodes describe a lamination step along one of the three orthogonal lamination directions, $\hat{x}$, $\hat{y}$, or $\hat{z}$. The volume fractions used in the different lamination steps are assigned to the edges of the tree, where $f_1$ is the volume fraction of phase $1$ in the final material. Thus, the volume fractions on the two edges entering each node sum to one.}
	\label{fig3}
\end{figure}

Hierarchical laminates are conveniently described using tree structures (see, e.g., Chap.\,9 in Ref.~\onlinecite{Milton:2002:TOC}). More precisely, every hierarchical laminate can be represented by a tree in which every node has either zero or two children. Nodes of the tree that have zero children, which are commonly referred to as leaves, correspond to one of the pure phases. Each node that is not a leaf, on the other hand, refers to a laminate that is formed from its children and, thus, has a certain lamination direction assigned to it. In the case of three-dimensional orthogonal laminates, there are only three possible lamination directions ($\hat{x}$, $\hat{y}$, and $\hat{z}$). Furthermore, we have to specify the volume fractions used in each lamination step. In our tree representation, we assign these volume fractions to the edges that connect the corresponding node to its children, i.e., the edges of the tree have certain weights. As an example, the tree structures of the laminates corresponding to the coated-cylinder assemblage and the coated-sphere assemblage are shown in Fig.\,\ref{fig3} (a) and (b), respectively.

\begin{figure}
	\includegraphics{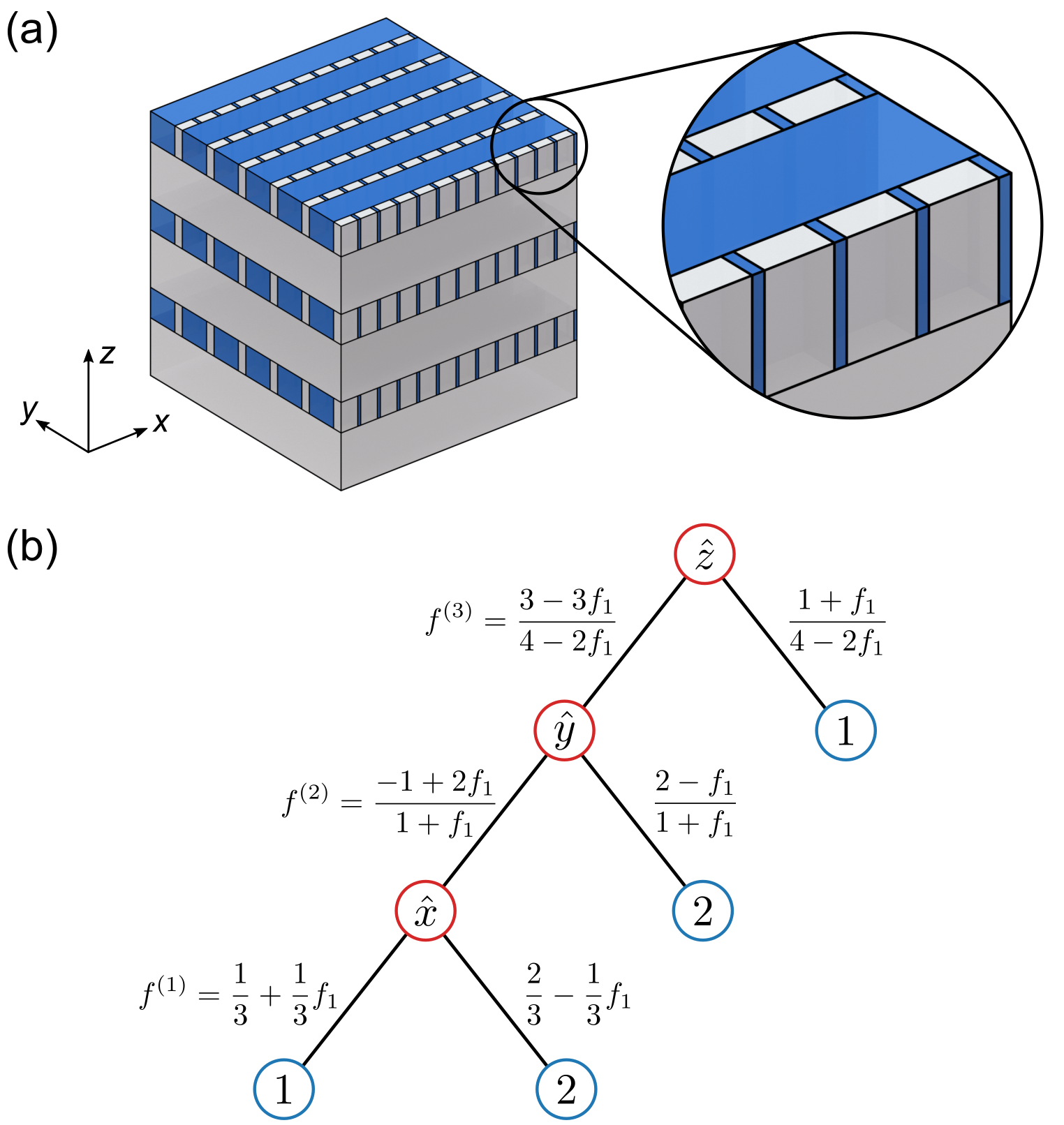}
	\caption{A schematic illustration (a) and the tree structure (b) of the optimal laminate (LA1) that is formed by laminating the rank-$2$ laminate corresponding to the coated-cylinder assemblage with phase $1$. The volume fractions used in the different lamination steps, $f^{(i)}$, are uniquely determined by the requirement that the structure has only one pole. The Schulgasser laminate
		formed from the anisotropic material obtained via this lamination scheme, i.e., from the material at the root of the tree, is an optimal isotropic material attaining the bound \eq{0.1}. In (a), phases $1$ and $2$ are shown in gray and blue, respectively.}
	\label{fig4}
\end{figure}

In order to identify additional microstructures attaining the bound \eq{0.1}, we now consider the class of hierarchical laminates that contains the five known optimal microstructures, i.e., the class of Schulgasser laminates formed from orthogonal laminates that have only one pole. More precisely, we form hierarchical laminates from laminates that are known to attain the bound (or one of the pure phases) in such a way that no additional poles are introduced. The first such hierarchical laminate (LA1) is formed from phase $1$ and the laminate corresponding to the coated-cylinder assemblage (with phase $1$ as the core phase and phase $2$ as the coating phase). It is closely related to the rank-$3$ coated laminate that corresponds to the coated-sphere assemblage. However, instead of laminating with phase $2$, i.e., the coating phase, in the third and last lamination step, we laminate with phase~$1$. An illustration of this hierarchical laminate as well as the corresponding tree structure are shown in Fig.\,\ref{fig4}. As indicated there, we denote the volume fractions used in the different lamination steps by $f^{(i)}$. These volume fractions are uniquely determined by the requirement that the pole that is formed in the last lamination step is identical to the pole of the laminate corresponding to the coated-cylinder assemblage, i.e.,
\beq \frac{\varepsilon_2}{\varepsilon_1}=\frac{f_2^{(2)}}{f_2^{(2)}-2}=\frac{\left(1-f_2^{(2)}\right)\left(f^{(3)}-1\right)-f^{(3)}}{\left(1-f^{(3)}\right)f_2^{(2)}},
\eeq{1.8}
where $f_2^{(2)}=1-f^{(1)}f^{(2)}$ is the volume fraction of phase~$2$ in the laminate corresponding to the coated-cylinder assemblage. As the volume fraction of phase $1$ in the final laminate is given by
\beq f_1=1-f^{(3)}f_2^{(2)},
\eeq{1.9}
and as all volume fractions have to lie in the range $\left[0,1\right]$, we can construct this laminate if only if ${f_1\in\left[1/2,1\right]}$. Analogously, we can construct the corresponding phase-interchanged hierarchical laminate (LA2) if and only if ${f_1\in\left[0, 1/2\right]}$. In the final step, in order to obtain isotropic optimal composites, we form Schulgasser laminates from the anisotropic hierarchical laminates LA1 and LA2. The $y$-transformed effective permittivities of these Schulgasser laminates are identical and given by the arithmetic mean of the $y$-transformed permittivities of the two Hashin-Shtrikman coated sphere assemblages,
\beq y_{\varepsilon}^{\text{LA1}}=y_{\varepsilon}^{\text{LA2}}=\frac{1}{2}\left(y_{\varepsilon}^{\text{CS}1}+y_{\varepsilon}^{\text{CS2}}\right)=\varepsilon_1+\varepsilon_2,
\eeq{1.10}
which implies that they attain the bound \eq{0.1}. 

\begin{figure}
	\includegraphics{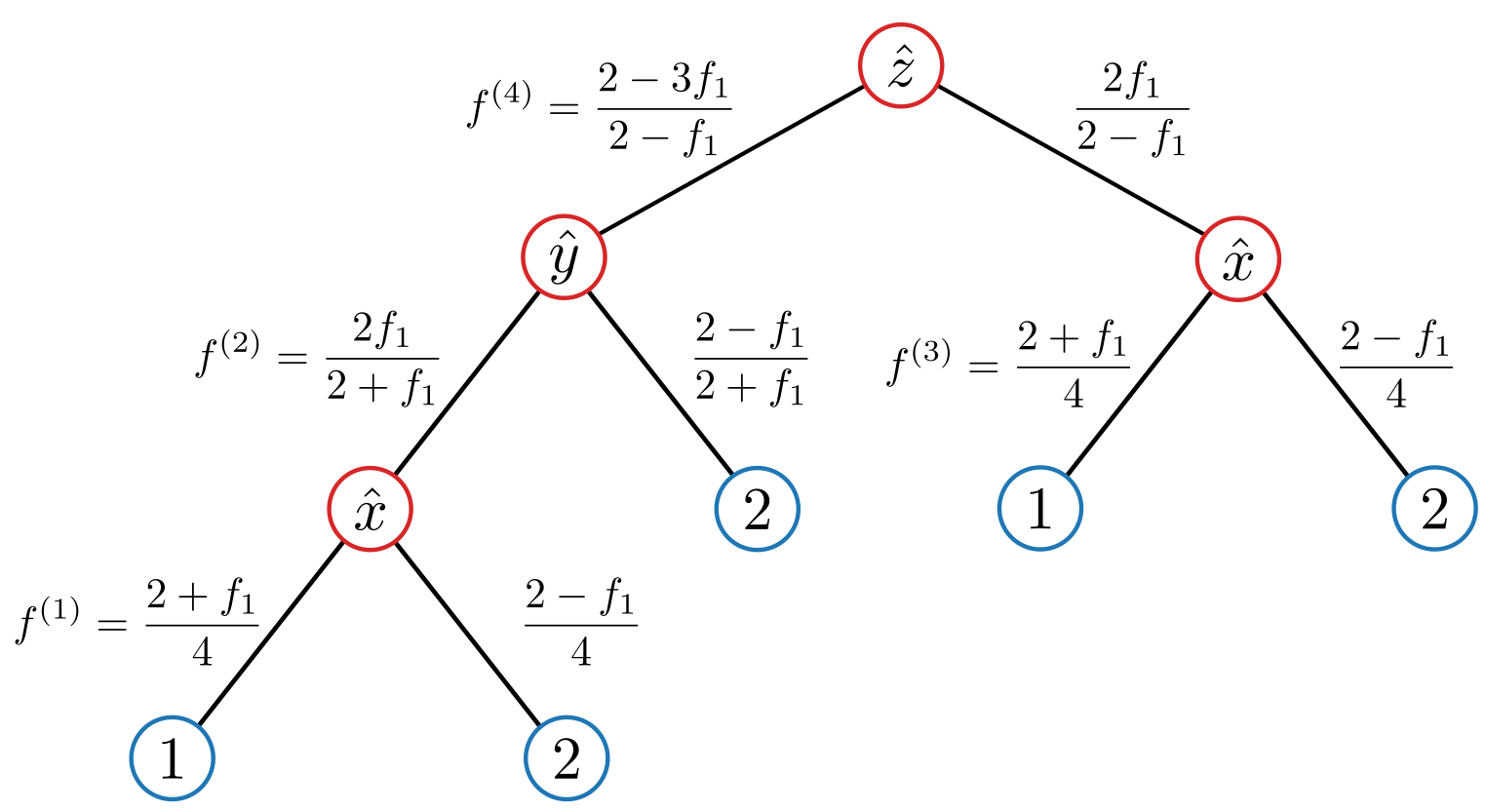}
	\caption{The tree structure of the optimal laminate (LB1) that is formed by laminating the rank-$2$ laminate corresponding to the coated-cylinder assemblage with a rank-$1$ laminate. The Schulgasser laminate made from this material attains the bound \eq{0.1}. As in the case of the laminate shown in Fig.\,\ref{fig4}, the volume fractions $f^{(i)}$ follow uniquely by requiring that the laminations do not lead to additional poles.}
	\label{fig5}
\end{figure}

In a similar fashion, we can obtain an additional optimal microstructure (LB1) by laminating the hierachical laminate corresponding to the coated-cylinder assemblage with a rank-$1$ laminate. The corresponding tree structure is shown in Fig.\,\ref{fig5}. Again, let $f^{(i)}$ denote the volume fractions used in the different lamination steps. We first require that the pole of the rank-$1$ laminate is identical to the pole of the laminate corresponding to the coated-cylinder assemblage. With $f^{(2)}_1=f^{(1)}f^{(2)}$ being the volume fraction of phase $2$ in the latter laminate, this condition reads
\beq \frac{\varepsilon_2}{\varepsilon_1}=\frac{f^{(3)}-1}{f^{(3)}}=\frac{f_1^{(2)}-1}{f_1^{(2)}+1}.
\eeq{1.11}
We then require that the pole created in the last lamination step is identical to the pole of the two laminates from which it is formed,
\beq \frac{\varepsilon_2}{\varepsilon_1}=\frac{f^{(3)}-1}{f^{(3)}}=\frac{\left(1-f^{(4)}\right)\left(f^{(3)}-1\right)+f^{(3)}}{\left(1-f^{(4)}\right)\left(f^{(3)}-1\right)+f^{(3)}-1}.
\eeq{1.12}
The volume fraction of phase $1$ in the final laminate is given by
\beq f_1=f^{(4)}f_1^{(2)}+\left(1-f^{(4)}\right)f^{(3)},
\eeq{1.13}
which, in combination with the conditions \eq{1.11} and \eq{1.12}, uniquely determines the volume fractions $f^{(i)}$. As these volume fractions have to lie in the range $\left[0,1\right]$, the laminate LB1 can be constructed if and only if $f_1\in \left[0,2/3\right]$. The $y$-transformed effective permittivity of the corresponding Schulgasser laminate is given by
\beq y_{\varepsilon}^{\text{LB1}}=f^{(4)}y_{\varepsilon}^{\text{CC1}}+\left(1-f^{(4)}\right)y_{\varepsilon}^{\text{L}}.
\eeq{1.14}
Similarly, we can find a corresponding laminate with interchanged phases (LB2), with
\beq y_{\varepsilon}^{\text{LB2}}=f^{(4)}y_{\varepsilon}^{\text{CC2}}+\left(1-f^{(4)}\right)y_{\varepsilon}^{\text{L}},
\eeq{1.15}
if and only if $f_1\in \left[1/3,0\right]$. As these points, $y_{\varepsilon}^{\text{LB1}}$ and $y_{\varepsilon}^{\text{LB2}}$, lie on the straight line joining $2\varepsilon_1$ and $2\varepsilon_2$, the Schulgasser laminates formed from these laminates attain the bound \eq{0.1}. 

\begin{figure*}
	\includegraphics{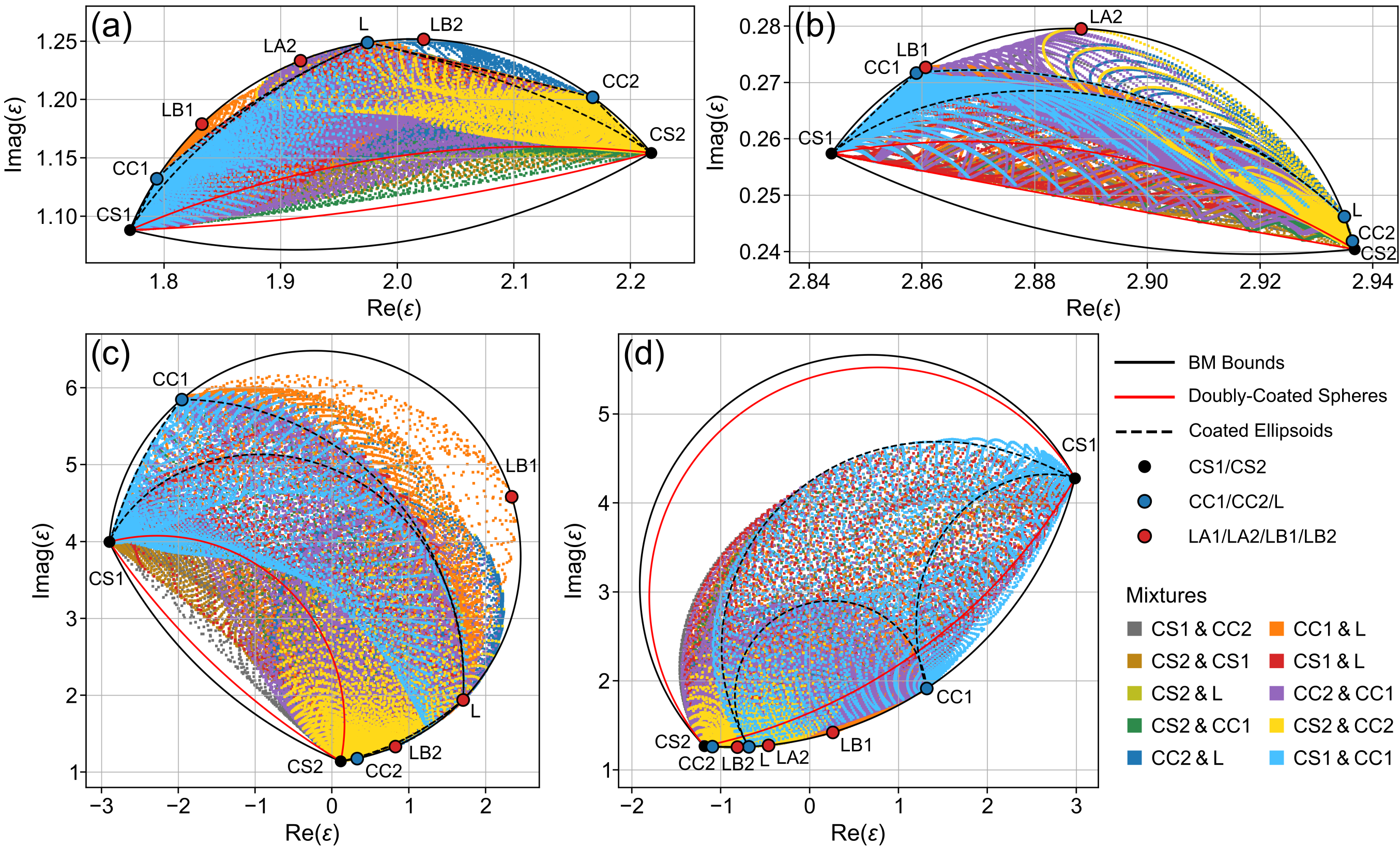}
	\caption{Numerical calculations of the effective complex permittivities (small squares) of Schulgasser laminates of microstructures formed by pairwise laminating the previously identified optimal laminates, CS1, CS2, CC1, CC2, and L. The different colors correspond to the different combinations of optimal laminates. The laminates formed in this way closely approach the BM bound corresponding to single-resonance structures, which is attained by the Hashin-Shtrikman coated sphere assemblages (black dots), the optimal microstructures previously identified by Milton (blue dots), and the hierarchical laminates described above (red dots). The boundary of the region attained by Schulgasser laminates of assemblages of coated ellipsoids is shown for reference (dashed curve). The two red curves correspond to assemblages of doubly-coated spheres. In Sec.\,III of this paper, we show that the outermost of these two curves is actually a bound. Note that the density of points in these plots in no way signifies the probability of obtaining such permittivities in experiments. The parameters are chosen as follows. In (a) and (b), corresponding to the situation far off resonance, the constituent permittivities are $\varepsilon_1=0.2+1.76\text{i}$ and $\varepsilon_2=3+0.1\text{i}$, respectively. While in (a) a moderate volume fraction of $f_1=0.4$ is chosen, (b) corresponds to the dilute limit, $f_1=0.05$. In both cases, for these specific choices of parameters, one of the hierarchical laminates identified here attains the largest possible imaginary part, i.e., it shows the strongest possible absorption. The parameters in (c) and (d) are $\varepsilon_1=-4.6+2.4\text{i}$, $\varepsilon_2=2.5+0.1\text{i}$, and $f_1=0.5$ and $\varepsilon_1=-4.6+2.4\text{i}$, $\varepsilon_2=2.5+0.1\text{i}$, and $f_1=0.4$, respectively.}
	\label{fig6}
\end{figure*}

Hence, in summary, we obtain three additional microstructures attaining the bound if $f_1\in (1/3,2/3)$ with $f_1\neq 1/2$ and two additional microstructures otherwise. While it remains presently unclear whether this strategy succeeds for every permittivity on the bound, using numerical calculations, one can interpolate between the known optimal laminates in such a way that the resulting laminates, which in general have more than a single pole, come very close to the bound. In order to do so, we choose two of the known optimal laminates (CS1, CS2, CC1, CC2, L), which we refer to as laminates A and B, and laminate them along one of the three axes in proportions $f$ and $1-f$. If $f_1^{\text{A}}$ and $f_1^{\text{B}}$ are the volume fractions of phase 1 in the laminates A and B, respectively, the volume fraction of phase 1 in the resulting material is given by 
\beq f_1=ff_1^{\text{A}}+(1-f)f_1^{\text{B}}.
\eeq{1.16}
As, in general, this laminate is not isotropic, we form a Schulgasser laminate in the final step. We then repeat this process for all three axes and for all possible pairwise combinations of the laminates CS1, CS2, CC1, CC2, and L and a large number of combinations of $f_1^{A}$ and $f_1^{B}$ while varying $f$ such that $f_1$ is kept fixed. We calculate the corresponding effective complex permittivities numerically. The results for different sets of parameters are shown in Fig.\,6. Both close to and far off resonance, the laminates closely approach the bound \eq{0.1}, which demonstrates that, for all practical purposes, it can be considered to be optimal. Moreover, the laminates fill the region between the bounds almost completely except for a gap close to the doubly-coated-sphere assemblage, which is especially pronounced in Fig.\,6(d). This gap can be readily filled by, e.g., forming a laminate with the doubly-coated-sphere assemblage.

\section{An optimal bound on the effective permittivity}
We now derive our improved bound on the isotropic effective electric permittivity that corresponds to the doubly-coated sphere assemblage. In the first step, following Cherkaev and Gibiansky \cite{Cherkaev:1994:VPC} (see also section\,11.5 in Ref.~\onlinecite{Milton:2002:TOC}), we rewrite the constitutive relation in terms of the real (primed) and imaginary (double-primed) parts of the fields,
\begin{eqnarray} &&\begin{pmatrix} \bm{e}'' \\ \bm{d}'' \end{pmatrix} = \bm{L} \begin{pmatrix} -\bm{d}' \\ \bm{e}' \end{pmatrix}\text{ with } \label{2.0}\\ 
&&\bm{L}= \begin{pmatrix} (\bm{\varepsilon}'')^{-1} & (\bm{\varepsilon}'')^{-1}\bm{\varepsilon}' \\ \bm{\varepsilon}'(\bm{\varepsilon}'')^{-1} &\bm{\varepsilon}'(\bm{\varepsilon}'')^{-1}\bm{\varepsilon}'+\bm{\varepsilon}''\end{pmatrix}\nonumber
\end{eqnarray}
being symmetric and positive definite. Now, we can recast the problem of finding the effective permittivity tensor as the following minimization variational principle (see also Ref.~\onlinecite{Cherkaev:1994:VPC}):
\begin{eqnarray} &&\begin{pmatrix} -\bm{d}'_0 \\ \bm{e}'_0 \end{pmatrix}\cdot\bm{L}_*\begin{pmatrix} -\bm{d}'_0 \\ \bm{e}'_0 \end{pmatrix} \label{2.1}\\
&&=\text{min}\left\lbrace\left\langle\begin{pmatrix} -\bm{d}' \\ \bm{e}' \end{pmatrix}\cdot\bm{L}\begin{pmatrix} -\bm{d}' \\ \bm{e}' \end{pmatrix}\right\rangle\left|\begin{array}{l}\langle\bm{d}'\rangle=\bm{d}'_0,~\nabla\cdot\bm{d}'=0\\\langle\bm{e}'\rangle=\bm{e}'_0,~\nabla\times\bm{e}'=0\end{array}\right.\right\rbrace. \nonumber
\end{eqnarray}
Using a constant trial field, one immediately obtains the arithmetic mean bound $\bm{L}_* \leq \langle \bm{L}\rangle$, while the corresponding dual variational principle
leads to the harmonic mean bound $\bm{L}^{-1}_* \leq \langle \bm{L}^{-1}\rangle$. Note that these two bounds are equivalent in the sense that they lead to the same bounds on the effective permittivity tensor \cite{Cherkaev:1994:VPC}. 

Before applying the translation method, we will first embed the variational problem \eq{2.1} in a variational problem involving a tensor of higher rank \cite{Tartar:1985:EFC, Avellaneda:1988:ECP, Milton:1990:BRT} (see also Sec.\,24.8 in Ref.~\onlinecite{Milton:2002:TOC}). Instead of working with the rank-2 tensor $\bm{L}$, we will consider a corresponding rank-4 tensor, $\bm{\mathcal{L}}$, that comprises several copies of $\bm{L}$. Intuitively speaking, this approach allows us to simultaneously probe the composite material along different directions. 

As the microscopic electric field, $\bm{e}(\bm{x})$, and the microscopic electric displacement field, $\bm{d}(\bm{x})$, depend linearly on the macroscopic electric field $\langle\bm{e}\rangle$ and $\langle \bm{d} \rangle$, we can write
\begin{eqnarray} 
\bm{e}(\bm{x})=\bm{E}(\bm{x})\langle\bm{e}\rangle \text{ and } \bm{d}(\bm{x})=\bm{D}(\bm{x})\langle\bm{e}\rangle, \label{2.102}
\end{eqnarray}
thereby introducing the rank-2 tensor fields $\bm{E}(\bm{x})$ and $\bm{D}(\bm{x})$, which, as opposed to a single solution of the quasistatic equations \eq{0.a}, fully characterize the composite material. The corresponding microscopic version of the constitutive law takes the form
\begin{eqnarray} 
\bm{D}=\bm{\mathcal{E}}:\bm{E},
\end{eqnarray}
where ``$:$'' denotes a contraction with respect to two indices, i.e., 
\begin{eqnarray} 
D_{ij}=\mathcal{E}_{ijkl}E_{kl},
\end{eqnarray}
and $\bm{\mathcal{E}}$ is a rank-$4$ tensor with components
\begin{eqnarray} 
\mathcal{E}_{ijkl} = \varepsilon_{ik}\delta_{jl} \text{ with } i,j,k,l \in \lbrace 1,2,3\rbrace.
\end{eqnarray}
The macroscopic version of the constitutive law, which defines the corresponding effective tensor, is given by
\beq \langle\bm{D}\rangle=\bm{\mathcal{E}}_*:\langle\bm{E}\rangle.
\eeq{2.3}
Using $\langle\bm{E}\rangle=\bm{I}$ and considering an isotropic composite, Eq.\,\eq{2.3} reduces to $\langle\bm{D}\rangle=\varepsilon_*\bm{I}$, where $\varepsilon_*$ is the scalar effective permittivity. 
As the fields $\bm{e}$ and $\bm{d}$ are curl- and divergence-free, respectively, we can find corresponding differential constraints on $\bm{E}$ and $\bm{D}$:
\beq \epsilon_{ijk}\partial_j E_{kl}=0 \text{ and } \partial_i D_{ij}=0.
\eeq{2.2}
As in the rank-2 case, cf. Eq.\,\eq{2.0}, we can rewrite the constitutive law in terms of the real and imaginary parts:
\begin{eqnarray} &&\begin{pmatrix} \bm{E}'' \\ \bm{D}'' \end{pmatrix} = \bm{\mathcal{L}}:\begin{pmatrix} -\bm{D}' \\ \bm{E}' \end{pmatrix} \text{ with }\\
&&\bm{\mathcal{L}}= \begin{pmatrix} (\bm{\mathcal{E}}'')^{-1} & (\bm{\mathcal{E}}'')^{-1}\bm{\mathcal{E}}' \\ \bm{\mathcal{E}}'(\bm{\mathcal{E}}'')^{-1} &\bm{\mathcal{E}}'(\bm{\mathcal{E}}'')^{-1}\bm{\mathcal{E}}'+\bm{\mathcal{E}}''\end{pmatrix},
\end{eqnarray}
where we have introduced the rank-4 tensor $\bm{\mathcal{L}}$.
The components of this tensor are related to the components of the corresponding rank-2 tensor as follows:
\beq\mathcal{L}_{ijkl}=L_{ik}\delta_{jl} \text{ with } j,l \in \lbrace 1,2,3\rbrace,~i,k \in \lbrace 1,\mydots,6\rbrace.
\eeq{2.2}
We now consider a material with translated properties,
\beq \widetilde{\bm{\mathcal{L}}}(\bm{x})=\bm{\mathcal{L}}(\bm{x})-\bm{T},
\eeq{2.10}
satisfying the following two conditions (see, e.g., Refs.~\onlinecite{Milton:1990:BRT,Gibiansky:1999:EPP} and Chap.\,24 in Ref.~\onlinecite{Milton:2002:TOC}):
\begin{enumerate}[label=(\roman*)]
\item The translated tensor is positive semidefinite, i.e., 
\beq \bm{\mathcal{L}}(\bm{x})-\bm{T} \geq 0,
\eeq{2.11}
which, as we are considering a two-phase medium, simplifies to 
\beq \bm{\mathcal{L}}_1-\bm{T} \geq 0 \text{ and } \bm{\mathcal{L}}_2-\bm{T} \geq 0.
\eeq{2.12}
\item The translation, $\bm{T}$, is quasiconvex, i.e.,
\beq \langle \bm{F}:\bm{T}:\bm{F}\rangle - \langle\bm{F}\rangle:\bm{T}:\langle\bm{F}\rangle \geq 0 
\eeq{2.13}
for all $\bm{F}=(-\bm{D}',\,\bm{E}')^{\intercal}$ with the periodic fields $\bm{D}'$ and $\bm{E}'$ subject to the usual differential constraints. 
\end{enumerate}
Note that if Eq.\,\eq{2.13} holds as an equality, $\bm{T}$ is said to be a null Lagrangian, or, more precisely, the quadratic form associated with $\bm{T}$ is a null Lagrangian.

Condition (i) allows us to apply the harmonic mean bound to the translated material,
\begin{eqnarray} \widetilde{\bm{\mathcal{L}}}_*^{-1}\leq \langle\widetilde{\bm{\mathcal{L}}}^{-1}\rangle,
\end{eqnarray}
and condition (ii) implies that
\begin{eqnarray} \widetilde{\bm{\mathcal{L}}}_*\leq \bm{\mathcal{L}}_*-\bm{T}.
\end{eqnarray}
In combination, we obtain the so-called translation bound:
\begin{eqnarray} \left(\bm{\mathcal{L}}_*-\bm{T}\right)^{-1} &&\leq \langle\left(\bm{\mathcal{L}}-\bm{T}\right)^{-1}\rangle \label{2.14}\\ &&=f_1\left(\bm{\mathcal{L}}_1-\bm{T}\right)^{-1}+f_2\left(\bm{\mathcal{L}}_2-\bm{T}\right)^{-1}.\nonumber
\end{eqnarray}
Analogous to the $y$-transform of the scalar effective parameters, Eq.\,\eq{0.6}, we can introduce the $y$-transform of the effective tensor $\bm{\mathcal{L}}_*$,
\begin{eqnarray} &&\bm{Y} = -f_1\bm{\mathcal{L}}_2-f_2\bm{\mathcal{L}}_1 \label{2.15}\\&&+f_1f_2(\bm{\mathcal{L}}_1-\bm{\mathcal{L}}_2)\cdot (f_1\bm{\mathcal{L}}_1+f_2\bm{\mathcal{L}}_2-\bm{\mathcal{L}}_*)^{-1}\cdot(\bm{\mathcal{L}}_1-\bm{\mathcal{L}}_2),\nonumber
\end{eqnarray}
in terms of which the bound takes the particularly simple form
\beq \bm{Y}+\bm{T}\geq 0.
\eeq{2.16}

While in principle, we could consider any arbitrary translation satisfying the aforementioned conditions, it has been shown that isotropic translations, reflecting the symmetry of the problem, are most-well suited \cite{Milton:1990:BRT}. We start by noting that any arbitrary isotropic rank-4 tensor can be written as 
\begin{eqnarray} &&A_{ijkl}(\lambda_1, \lambda_2, \lambda_3) = \frac{\lambda_1}{3}\delta_{ij}\delta_{kl} \label{2.17}\\
&&+\frac{\lambda_2}{2}\left(\delta_{ik}\delta_{jl}+\delta_{il}\delta_{jk}-\frac{2}{3}\delta_{ij}\delta_{kl}\right)+\frac{\lambda_3}{2}\left(\delta_{ik}\delta_{jl}-\delta_{il}\delta_{jk}\right),\nonumber
\end{eqnarray}
where the three terms correspond to projections of an arbitrary rank-2 tensor onto the subspaces of tensors proportional to the identity tensor, trace-free symmetric tensors, and antisymmetric tensors. We refer to these three terms as the bulk-modulus, shear-modulus, and antisymmetrization terms, since in elasticity the coefficients $\lambda_1$ and $\lambda_2$ correspond to the bulk and shear modulus, respectively, while $\lambda_3$ has no counterpart as the elasticity tensor is symmetric with respect to permutations of the first two (as well as the last two) indices. In the following, we encounter tensors of the form
\beq \begin{pmatrix} \bm{A}(\lambda_1, \lambda_2, \lambda_3) & \bm{A}(\lambda_4, \lambda_5, \lambda_6) \\ \bm{A}(\lambda_4, \lambda_5, \lambda_6) & \bm{A}(\lambda_7, \lambda_8, \lambda_9) \end{pmatrix}.
\eeq{2.18}
We use the fact that such a tensor is positive semidefinite if and only if the three matrices,
\beq \begin{pmatrix} \lambda_1 & \lambda_4 \\ \lambda_4 & \lambda_7 \end{pmatrix},~\begin{pmatrix} \lambda_2 & \lambda_5 \\ \lambda_5 & \lambda_8 \end{pmatrix}, \text{ and }\begin{pmatrix} \lambda_3 & \lambda_6 \\ \lambda_6 & \lambda_9 \end{pmatrix}
\eeq{2.19}
are positive semidefinite. We now chose the translation as
\beq \bm{T}(t_1,t_2,t_3) = \begin{pmatrix} \bm{A}(-t_1, 2t_1, 0) & \bm{A}(-t_3, -t_3, -t_3) \\ \bm{A}(-t_3, -t_3, -t_3) & \bm{A}(-2t_2, t_2, -t_2) \end{pmatrix},
\eeq{2.20}
for $t_1\geq 0$ and $t_2$ and $t_3$ arbitrary.

In order to show that such a translation is quasiconvex, as shown by Tartar \cite{Tartar:1979:CCA, Tartar:1985:EFC} and Murat and Tartar \cite{Murat:1985:CVH} (see also section\,24.3 in Ref.~\onlinecite{Milton:2002:TOC}), and following the analogous discussion in Ref.~\onlinecite{Gibiansky:1993:EVM}, it turns out to be useful to consider the Fourier series of the real parts the fields,
\begin{eqnarray}
\bm{E'}&&=\langle\bm{E'}\rangle+\sum_{k\neq 0}e^{-i\bm{k}\cdot\bm{x}}\bm{E}_{\bm{k}}, \label{2.210} \\
\text{and }\bm{D'}&&=\langle\bm{D'}\rangle+\sum_{k\neq 0}e^{-i\bm{k}\cdot\bm{x}}\bm{D}_{\bm{k}}. \label{2.211}
\end{eqnarray}
Since $\bm{E}$ and $\bm{D}$ are curl- and divergence-free, respectively, we obtain the following constraints on the Fourier coefficients
\beq \bm{k}\times\bm{E}_{\bm{k}}=0 \text{ and } \bm{k}\cdot\bm{D}_{\bm{k}}=0.
\eeq{2.212}

Using Plancherel's identity, one finds that, in order to prove the quasiconvexity of the translation, it suffices to show that
\begin{eqnarray}
&& \sum_{\bm{k}\neq 0}\overbar{\bm{D}}_{\bm{k}}:\bm{A}(-t_1,2t_1,0):\bm{D}_{\bm{k}}\geq 0, \label{2.213}\\
&& \sum_{\bm{k}\neq 0}\overbar{\bm{E}}_{\bm{k}}:\bm{A}(-2t_2,t_2,-t_2):\bm{E}_{\bm{k}}=0, \label{2.214}\\
\text{and }&& \sum_{\bm{k}\neq 0}\overbar{\bm{D}}_{\bm{k}}:\bm{A}(-t_3,-t_3,-t_3):\bm{E}_{\bm{k}}=0, \label{2.215}
\end{eqnarray}
where $\overbar{\bm{E}}_{\bm{k}}$ denotes the complex conjugate of $\bm{E}_{\bm{k}}$. Note that we choose $\bm{A}(-2t_2,t_2,-t_2)$ and $\bm{A}(-t_3,-t_3,-t_3)$ to be null Lagrangians (with respect to the appropriate subspaces) rather than merely satisfying the condition of quasiconvexity. As we are considering isotropic tensors, it is sufficient to consider a single choice of $\bm{k}$. For example, for $\bm{k}=(1,0,0)^{\intercal}$ the constraints \eq{2.212} imply that the fields have the form
\beq \bm{D}_{\bm{k}}=\begin{pmatrix} 0 & 0 & 0 \\
D_{21} & D_{22} & D_{23}\\
D_{31} & D_{32} & D_{33} \end{pmatrix} \text{ and } \bm{E}_{\bm{k}}=\begin{pmatrix} E_{11} & E_{12} & E_{13} \\
0 & 0 & 0\\
0 & 0 & 0 \end{pmatrix},
\eeq{2.2}
and it becomes straightforward to show that the inequalities \eq{2.213}-\eq{2.215} hold.

Having established that a translation of the form given in Eq.\,\eq{2.20} is quasiconvex, we can now return to translation bound. Introducing the $y$-transform of the effective tensor $\bm{\mathcal{E}}_*$,
\begin{eqnarray} &&\bm{\mathcal{Y}} = -f_1\bm{\mathcal{E}}_2-f_2\bm{\mathcal{E}}_1\\&&+f_1f_2(\bm{\mathcal{E}}_1-\bm{\mathcal{E}}_2)\cdot (f_1\bm{\mathcal{E}}_1+f_2\bm{\mathcal{E}}_2-\bm{\mathcal{E}}_*)^{-1}\cdot(\bm{\mathcal{E}}_1-\bm{\mathcal{E}}_2),\nonumber
\end{eqnarray}
we can write (see, e.g., Ref.~\onlinecite{Gibiansky:1993:EVM})
\begin{eqnarray} \bm{Y}= \begin{pmatrix} (\bm{\mathcal{Y}}'')^{-1} & -(\bm{\mathcal{Y}}'')^{-1}\bm{\mathcal{Y}}' \\ -\bm{\mathcal{Y}}'(\bm{\mathcal{Y}}'')^{-1} &\bm{\mathcal{Y}}'(\bm{\mathcal{Y}}'')^{-1}\bm{\mathcal{Y}}'+\bm{\mathcal{Y}}''\end{pmatrix}.
\end{eqnarray}
Restricting ourselves to isotropic effective permittivities,
\beq \bm{\mathcal{Y}}=y_{\varepsilon}\bm{A}(1,1,1),
\eeq{2.2}
we then obtain
\begin{eqnarray} &&\bm{Y}=\begin{pmatrix} (y_{\varepsilon}'')^{-1}\bm{A}(1, 1, 1) & -(y_{\varepsilon}'')^{-1}y_{\varepsilon}'\bm{A}(1, 1, 1) \\ -(y_{\varepsilon}'')^{-1}y_{\varepsilon}'\bm{A}(1, 1, 1) & ((y_{\varepsilon}'')^{-1}(y_{\varepsilon}')^2+y_{\varepsilon}'')\bm{A}(1, 1, 1) \end{pmatrix}.\nonumber\\
&&
\end{eqnarray}
Using the decomposition into the bulk-modulus, shear-modulus, and antisymmetrization terms, it becomes clear that the translation bound reduces to the three conditions
\begin{eqnarray} &&\begin{pmatrix} (y_{\varepsilon}'')^{-1}-t_1 & -(y_{\varepsilon}'')^{-1}y_{\varepsilon}'-t_3 \\ -(y_{\varepsilon}'')^{-1}y_{\varepsilon}'-t_3 & (y_{\varepsilon}'')^{-1}(y_{\varepsilon}')^2+y_{\varepsilon}''-2t_2 \end{pmatrix}\geq 0, \label{2.30}\\
&&\begin{pmatrix} (y_{\varepsilon}'')^{-1}+2t_1 & -(y_{\varepsilon}'')^{-1}y_{\varepsilon}'-t_3 \\ -(y_{\varepsilon}'')^{-1}y_{\varepsilon}'-t_3 & (y_{\varepsilon}'')^{-1}(y_{\varepsilon}')^2+y_{\varepsilon}''+t_2 \end{pmatrix}\geq 0, \label{2.31}\\
&&\begin{pmatrix} (y_{\varepsilon}'')^{-1} & -(y_{\varepsilon}'')^{-1}y_{\varepsilon}'-t_3 \\ -(y_{\varepsilon}'')^{-1}y_{\varepsilon}'-t_3 & (y_{\varepsilon}'')^{-1}(y_{\varepsilon}')^2+y_{\varepsilon}''-t_2 \end{pmatrix}\geq 0, \label{2.32}
\end{eqnarray}
which imply that the corresponding determinants are non-negative. Thus, we obtain from the first of these conditions, Eq.\,\eq{2.30},
\beq
\frac{t_1}{y_{\varepsilon}''}\left((y_{\varepsilon}'-y_{\text{c}}')^2+(y_{\varepsilon}''-y_{\text{c}}'')^2-R^2\right)\leq 0,
\eeq{2.33}
where we have introduced the parameters
\begin{eqnarray}
&&y'_{\text{c}} = -\frac{t_3}{t_1},~ y''_{\text{c}} = \frac{1+2t_1t_2-t_3^2}{2t_1}, \label{2.34}\\ 
&&\text{ and } R=\left|\frac{1-2t_1t_2+t_3^2}{2t_1}\right|. \nonumber
\end{eqnarray}
Hence, the $y$-transformed effective permittivity has to lie inside of or on a circle with center $(y'_{\text{c}},y''_{\text{c}})$ and radius $R$. Choosing different translations, i.e., different values of $t_1$, $t_2$, and $t_3$, corresponds to moving and scaling the circle in the complex plane. In contrast to the complex bulk modulus case \cite{Gibiansky:1993:EVM}, the circle may or may not contain the origin, as 
\beq (y'_{\text{c}})^2+(y''_{\text{c}})^2-R^2=2\frac{t_2}{t_1}
\eeq{2.35}
is not necessarily non-negative. As shown below, the first condition leads to bounds that constrain the $y$-transformed effective permittivity to a region in the complex plane that is bounded by a circular arc and a straight line. As the circular arc turns out to correspond to the doubly-coated-sphere assemblage and the straight line corresponds to the bound \eq{0.1}, which is the tightest possible bound that is volume fraction independent in the $y$-plane, the second and third condition cannot provide any additional information and may be disregarded. 

We now identify the restrictions on the parameters $y'_{\text{c}}$, $y''_{\text{c}}$, and $R$, i.e., on the choice of circles bounding the $y$-transformed effective permittivity, imposed by the condition that the translated tensor is positive semidefinite, i.e., by Eq.\,\eq{2.11}. Using the fact that the permittivity tensors of the two phases are isotropic,
\beq \bm{\mathcal{E}}_i=\varepsilon_i\bm{A}(1,1,1) \text{ for } i \in \lbrace 1,2\rbrace,
\eeq{2.37}
we find that
\beq \bm{\mathcal{L}}_i=\begin{pmatrix} (\varepsilon''_i)^{-1}\bm{A}(1, 1, 1) & (\varepsilon''_i)^{-1}\varepsilon'_i\bm{A}(1, 1, 1) \\ (\varepsilon''_i)^{-1}\varepsilon'_i\bm{A}(1, 1, 1) & ((\varepsilon''_i)^{-1}(\varepsilon'_i)^2+\varepsilon''_i)\bm{A}(1, 1, 1) \end{pmatrix}.
\eeq{2.38}
Again using the decomposition into the bulk-modulus, shear-modulus, and antisymmetrization terms, we find that the positive-semidefiniteness of the translated tensor is equivalent to the three constraints
\begin{eqnarray} &&\begin{pmatrix} (\varepsilon''_i)^{-1}+t_1 & (\varepsilon''_i)^{-1}\varepsilon'_i+t_3 \\ (\varepsilon''_i)^{-1}\varepsilon'_i+t_3 & (\varepsilon''_i)^{-1}(\varepsilon'_i)^2+\varepsilon''_i+2t_2 \end{pmatrix}\geq 0,\label{2.39}\\
&&\begin{pmatrix} (\varepsilon''_i)^{-1}-2t_1 & (\varepsilon''_i)^{-1}\varepsilon'_i+t_3 \\ (\varepsilon''_i)^{-1}\varepsilon'_i+t_3 & (\varepsilon''_i)^{-1}(\varepsilon'_i)^2+\varepsilon''_i-t_2 \end{pmatrix}\geq 0,\label{2.4001}\\
&&\begin{pmatrix} (\varepsilon''_i)^{-1} & (\varepsilon''_i)^{-1}\varepsilon'_i+t_3 \\ (\varepsilon''_i)^{-1}\varepsilon'_i+t_3 & (\varepsilon''_i)^{-1}(\varepsilon'_i)^2+\varepsilon''_i+t_2 \end{pmatrix}\geq 0,\label{2.401}
\end{eqnarray}
which imply that the corresponding determinants are non-negative. Evaluating the first two determinants gives
\begin{eqnarray} &&(y_{\text{c}}'+\varepsilon_i')^2+(y_{\text{c}}''+\varepsilon_i'')^2\geq R^2 \label{2.402}\\
\text{and }&&(y_{\text{c}}'-2\varepsilon_i')^2+(y_{\text{c}}''-2\varepsilon_i'')^2\leq R^2. \label{2.403}
\end{eqnarray}
By considering the remaining principal minors, i.e., the diagonal elements, it can be shown that Eq.\,\eq{2.402} and Eq.\,\eq{2.403} are not only necessary but also sufficient for  the first two constraints, Eq.\,\eq{2.39} and Eq.\,\eq{2.4001}. Furthermore, the third constraint, Eq.\,\eq{2.401}, can be discarded, as it can be written as a weighted arithmetic mean of the other two constraints. 
\begin{figure}
	\includegraphics{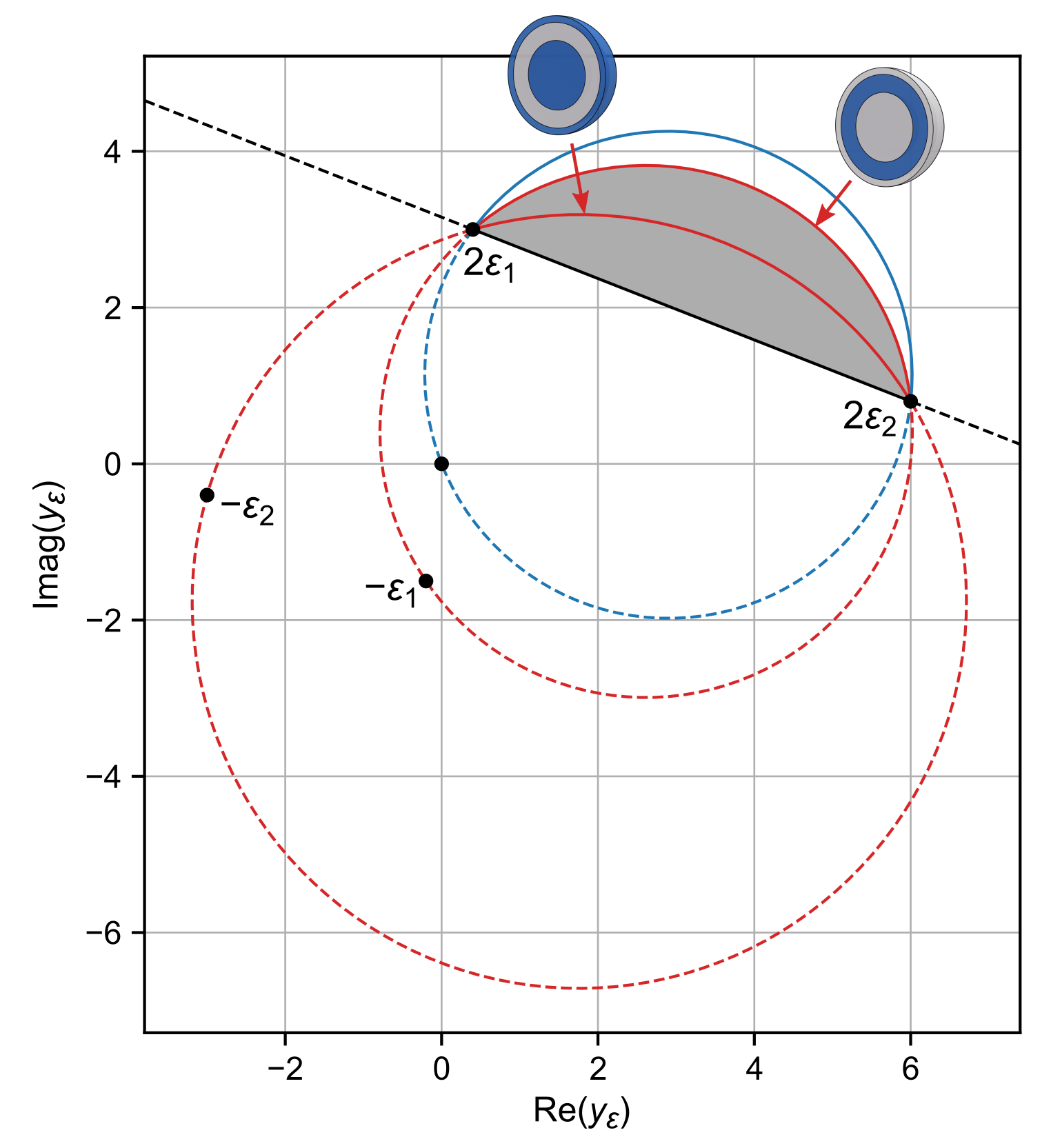}
	\caption{An illustration of the bounds on the $y$-transformed effective electric permittivity. The analytic bounds by Bergman and Milton correspond to the straight black line joining $2\varepsilon_1$ and $2\varepsilon_2$, and the blue circular arc passing through $2\varepsilon_1$, $2\varepsilon_2$, and the origin. Using variational methods, we find that $y$-transformed effective electric permittivity is confined to the gray-shaded region, i.e., the region that is additionally bounded by outermost of the two red circular arcs passing through $2\varepsilon_1$, $2\varepsilon_2$, and $-\varepsilon_1$ or $-\varepsilon_2$, which correspond to Hashin-Shtrikman assemblages of doubly-coated spheres. The parameters are $\varepsilon_1=0.2+1.5\text{i}$ and $\varepsilon_2=3+0.4\text{i}$.}
	\label{fignewbound}
\end{figure}

Hence, the restriction to positive-semidefinite translated tensors corresponds to choosing the parameters of the translation such that $2\varepsilon_1$ and $2\varepsilon_2$ do not lie outside of the circle and $-\varepsilon_1$ and $-\varepsilon_2$ do not lie inside of the circle. The extremal translations consistent with this restriction correspond to a generalized circle, the half-space bounded by the straight line between $2\varepsilon_1$ and $2\varepsilon_2$ that does not contain $-\varepsilon_1$ and $-\varepsilon_2$, and one of the two circles passing through $2\varepsilon_1$, $2\varepsilon_2$, and $-\varepsilon_1$ or $-\varepsilon_2$. 

Thus, we find that, as illustrated in Fig.\,\ref{fignewbound}, the $y$-transformed effective complex electric permittivity is bounded by the straight line joining $2\varepsilon_1$ and $2\varepsilon_2$ on one side and the outermost of the two circular arcs passing through $2\varepsilon_1$, $2\varepsilon_2$ and $-\varepsilon_1$ or $-\varepsilon_2$ on the other side. While the former bound, which is the one given in Eq.\,\eq{0.1}, has been previously derived using the analytic method, the latter bound is tighter than any previously identified bound and even turns out to be optimal, as it corresponds to assemblages of doubly-coated spheres. The effective permittivity of such an assemblage (see, e.g., Sec.\,7.2 in Ref.~\onlinecite{Milton:2002:TOC} for a detailed discussion), with phase $1$ in the core and the outer shell and phase $2$ in the inner shell, is given by
\beq y_{\varepsilon}^{\text{DCS1}}=2\varepsilon_1\frac{3p\varepsilon_2+(1-p)(\varepsilon_1+2\varepsilon_2)}{3p\varepsilon_1+(1-p)(\varepsilon_1+2\varepsilon_2)},
\eeq{2.37}
where the volume fractions of the core, the inner shell, and the outer shell are $p f_1$, $1-f_1$, and $(1-p)f_1$, respectively. Clearly, this equation corresponds to a circular arc passing through $2\varepsilon_1$, $2\varepsilon_2$, and $-\varepsilon_1$. For the phase-interchanged case, i.e., phase $2$ in the core and the outer shell and phase $1$ in the inner shell, one obtains
\beq y_{\varepsilon}^{\text{DCS2}}=2\varepsilon_2\frac{3p\varepsilon_1+(1-p)(\varepsilon_2+2\varepsilon_1)}{3p\varepsilon_2+(1-p)(\varepsilon_2+2\varepsilon_1)},
\eeq{2.37}
which corresponds to a circular arc passing through $2\varepsilon_1$, $2\varepsilon_2$, and $-\varepsilon_2$. Here, the volume fractions of the core, the inner shell, and the outer shell are $p (1-f_1)$, $f_1$, and $(1-p)(1-f_1)$, respectively.

\section{Relation to bounds on the complex polarizability}
As shown in Ref.~\onlinecite{Milton:1981:TPA}, bounds on the effective complex permittivity for small values of $f_1$ directly lead to bounds on
the orientation averaged complex polarizability of an inclusion (or set of inclusions) having permittivity $\Gve_1$ embedded in a medium with permittivity
$\Gve_2$. Independently, Miller and coauthors \cite{Miller:2014:FLE} have derived explicit bounds on the imaginary part of the polarizability (which describes the absorption of electromagnetic
radiation by a cloud of small particles, each much smaller than the wavelength) and subsequently, in Ref.~\onlinecite{Milton:2017:BCP}, explicit bounds have been obtained on the
complex polarizabilty (not just its imaginary part) by taking the dilute limit $f_1\to 0$ in the BM bounds, keeping the leading term in $f_1$. It is
important to remark that the elementary arguments of Ref.~\onlinecite{Miller:2016:FLO} and its recent generalizations incorporating size-dependent radiation effects \cite{Gustafsson:2020:BAS, Kuang:2020:MER, Molesky:2020:TES} lead to useful bounds valid at any wavelength, not necessarily large compared to the particle size. Our improved bound naturally also applies when the volume fraction $f_1$ is small and thus produces a tighter and optimal bound on the complex polarizability. 

Consider, in the quasistatic regime, a dilute suspension of randomly oriented identical particles in vacuum or air. The effective relative permittivity of such a cloud of particles is given by
\beq \bm{\varepsilon_*}\approx1+f_1\frac{\text{Tr}(\bm{\alpha})}{3V},
\eeq{2.40}
where $\text{Tr}(\bm{\alpha})/3$ and $V$ are the angle-averaged polarizability tensor and the volume of each of the particles. Then, it follows from the bound \eq{0.1} and our new bound corresponding to the doubly-coated-sphere assemblage that the orientation-averaged polarizability per unit volume, $\text{Tr}(\bm{\alpha})/(3V)$, is bounded by the circular arc
\beq \alpha^{\text{BM}}(u)=\chi_1-\frac{\chi_1^2}{1+\chi_1+2(u\chi_1+1)}
\eeq{2.41}
and the outermost of the straight line-segment
\beq \alpha^{\text{DCS1}}(u)=\frac{3\chi_1}{\chi_1+3}+\frac{2u\chi_1^3}{3(\chi_1+1)(\chi_1+3)}
\eeq{2.42}
and the circular arc
\beq \alpha^{\text{DCS2}}(u)=\frac{3\chi_1(3+2\chi_1)}{9(\chi_1+1)+2u\chi_1^2},
\eeq{2.43}
where $\chi_1$ is the electric susceptibility of the constituent material of the particles and $u \in [0,1]$. This result improves on the bounds derived in Ref.~\onlinecite{Milton:2017:BCP} that follow from the BM bounds.

We immediately obtain a corresponding bound on the angle-averaged extinction cross-section per unit volume of a quasistatic particle \cite{Bohren:1988:ASL},
\beq \frac{\sigma_{\text{ext}}}{V}=\frac{2\pi}{\lambda}\text{Im}\left(\frac{\text{Tr}(\bm{\alpha})}{3V}\right),
\eeq{2.44}
which is a measure of the efficiency at which a particle scatters and absorbs light. 
This new bound reads as
\begin{eqnarray} 
\frac{\sigma_{\text{ext}}}{V}\leq\frac{2\pi}{\lambda}\underset{0\leq u \leq 1}{\text{max}}\left\lbrace\text{Im}\left(\alpha^{\text{BM}}(u)\right),\,\text{Im}\left(\alpha^{\text{DCS2}}(u)\right)\right\rbrace,
\end{eqnarray}
and improves on the bounds by Miller \textit{et al}. \cite{Miller:2014:FLE}, which follow from the BM bounds. The (albeit typically small) improvement over these previously derived bounds is obtained if the maximum corresponds to a point on the arc $\alpha^{\text{DCS2}}(u)$ with $u \in (0,1)$. For example, this is the case for materials with a large negative real part of the susceptibility, i.e., metals, which give the largest values of the extinction cross-section per unit volume. Ideally, to maximize the absorption, one chooses a metal with a small imaginary part of the susceptibility.

\section{Summary and Conclusions}
We studied the range of effective complex permittivities of a three-dimensional isotropic composite material made from two isotropic phases. We started from the well-known Bergman-Milton bounds, which bound the effective complex permittivity by two circular arcs in the complex plane. In the first step, we showed that several points on one of these arcs are attained by a specific class of hierarchical laminates. Furthermore, on the basis of numerical calculations, we showed that there is a natural way of interpolating between these laminates, which results in laminates approaching the arc in the gaps between these points. We then showed, using established variational methods, that the second arc can be replaced by an optimal bound that corresponds to assemblages of doubly coated spheres. Using this result, we derived corresponding bounds on the angle-averaged polarizability and the extinction cross-section of small particles. While we have focused on bounds using the quasistatic approximation, our results should be a useful benchmark for future bounds that might be more generally valid. For example, bounds on the absorption and scattering of radiation by particles of general shape and valid at any frequency have been derived in Ref.\,\onlinecite{Miller:2016:FLO}. While these bounds are quite tight, they are not as tight as our bounds in the quasistatic limit. This shows that there is room for improvement, perhaps using some sort of hybrid bounding method.

\section*{Acknowledgments}
C.K. and G.W.M. are grateful to the National Science Foundation for support through the Research Grant No. DMS-1814854. O.D.M. was supported by the Air Force Office of Scientific Research under Award No. FA9550-17-1-0093. O.D.M. and G.W.M. thank the Korean Advanced Institute of Science and Technology (KAIST) mathematical research station (KMRS), where this work was initiated, for support through a chair professorship to G.W.M. We also thank Aaron Welters (Florida Institute of Technology) for providing some additional early references to the BM bounds in the wider context of Stieltjes functions.

\end{document}